\documentclass{sig-alternate-05-2015}

\usepackage{comment}
\usepackage{pbox}
\usepackage{graphicx}
\usepackage{mathptmx} 
\usepackage{helvet} 
\usepackage{courier}
\usepackage{type1cm} 
\usepackage[table]{xcolor}
\usepackage{makeidx}
\usepackage{graphicx}
\usepackage{multicol} 
\usepackage[bottom]{footmisc}
\usepackage{multirow}
\usepackage{epsfig}
\usepackage{rotating}
\usepackage{lscape}
\usepackage{algorithm}
\usepackage{algpseudocode}
\usepackage{pifont}
\usepackage{cite}

\begin{document}

\setcopyright{acmcopyright}
\title{IGLOO: Integrating global and local biological network alignment}
\numberofauthors{4} 
\author{
\alignauthor
Lei Meng\\
\affaddr{Department of Computer Science and Engineering,}\\
\affaddr{ECK Institute for Global \\Health, Interdisciplinary Center for Network \\Science and Applications,}\\
\affaddr{and Wireless Institute}\\
\affaddr{University of Notre Dame}
\alignauthor
Joseph Crawford\\
\affaddr{Department of Computer Science and Engineering,}\\
\affaddr{ECK Institute for Global \\Health, and Interdisciplinary Center for Network \\Science and Applications,}\\
\affaddr{University of Notre Dame}
\alignauthor Aaron Striegel\\
\affaddr{Department of Computer Science and Engineering}\\
\affaddr{and Wireless Institute}\\
\affaddr{University of Notre Dame}
\and 
\alignauthor Tijana Milenkovi\'{c}\titlenote{Corresponding author (e-mail: tmilenko@nd.edu)}\\
\affaddr{Department of Computer Science and Engineering,}\\
\affaddr{ECK Institute for Global \\Health, and Interdisciplinary Center for Network \\Science and Applications,}\\
\affaddr{University of Notre Dame}
}
\maketitle
\label{sec:new_method}

\begin{abstract}
Analogous to genomic sequence alignment, biological network ali-
gnment 
(NA) aims to find regions of similarities between molecular networks
(rather than sequences) of different species. NA can be either local
(LNA) or global (GNA). LNA aims to identify highly conserved common
subnetworks, which are typically small, while GNA aims to identify
large common subnetworks, which are typically suboptimally
conserved. We recently showed that LNA and GNA yield complementary
results: LNA has high functional but low topological alignment
quality, while GNA has high topological but low functional alignment
quality. Thus, we propose IGLOO, a new approach that integrates GNA
and LNA in hope to reconcile the two. We evaluate IGLOO against
state-of-the-art LNA (NetworkBLAST, NetAligner, AlignNemo, and
AlignMCL) and GNA (GHOST, NETAL, GEDEVO, MAGNA++, WAVE, and
L-GRAAL) methods. We show that IGLOO allows for a trade-off between
topological and functional alignment quality better than the existing
LNA and GNA methods considered in our study.
\end{abstract}

\section{Introduction}

Large amounts of protein-protein interaction (PPI) data have become
available due to the advancement of high throughput biotechnologies
for data collection \cite{I2D2007,BIOGRID}. In PPI networks, nodes are
proteins and edges correspond to physical interactions between the
proteins. Network alignment (NA) of PPI data across species is gaining
importance. This is because NA aims to find a good node mapping between PPI networks of different
species that identifies topologically and functionally similar (i.e.,
conserved) network regions \cite{Fazle2015}, and as such, it can be used to transfer biological knowledge from
well- to poorly-studied species between such conserved network regions.  Consequently, NA is expected to lead to new discoveries in evolutionary biology. We note that while we
focus on NA in the domain of computational biology, NA has
applications in other domains, such as online social networks
\cite{Narayanan2011}, pattern recognition
\cite{pm1,pm2}, and language processing \cite{symword}. Our work is applicable to any of these domains.

NA is computationally
intractable, since the underlying subgraph isomorphism problem, which
determines if a network is an exact subgraph of another network, is
NP-complete. Therefore, efficient heuristic approaches are needed to
solve the NA problem approximately.

\begin{figure}
\centering
\textbf{(a)}\includegraphics[width=.4\columnwidth]{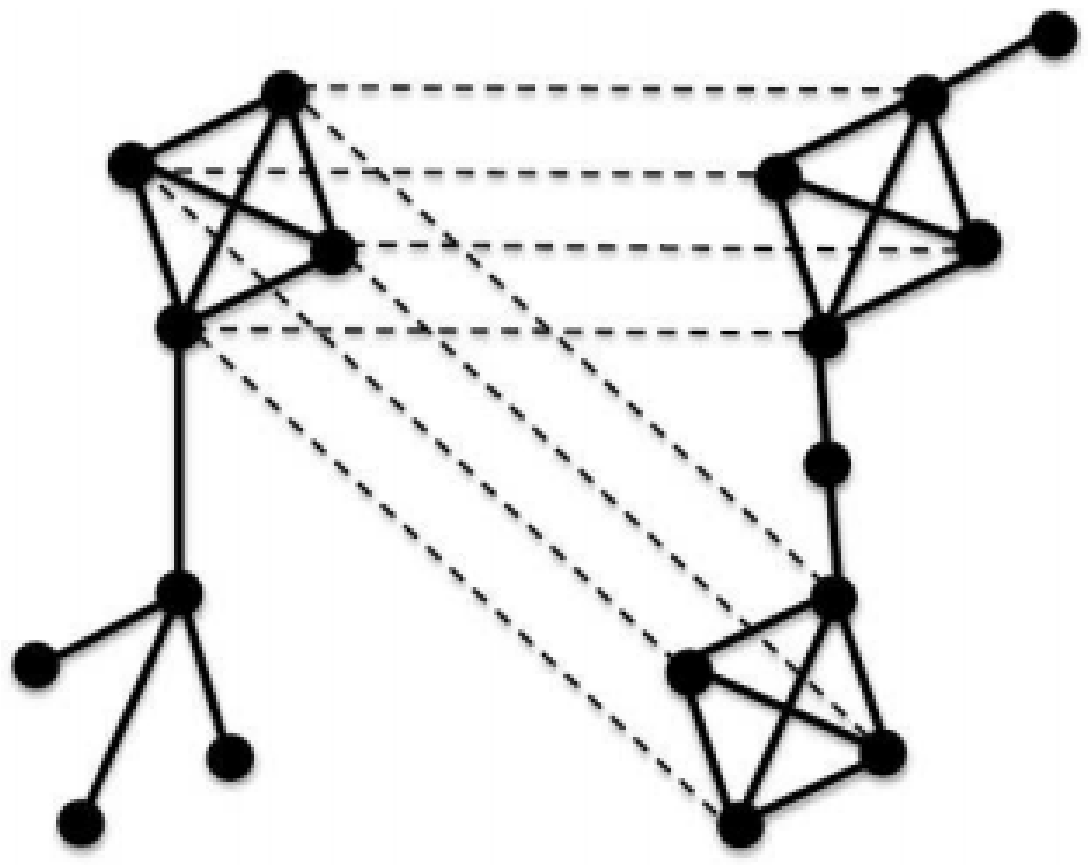}
\textbf{(b)}\includegraphics[width=.37\columnwidth]{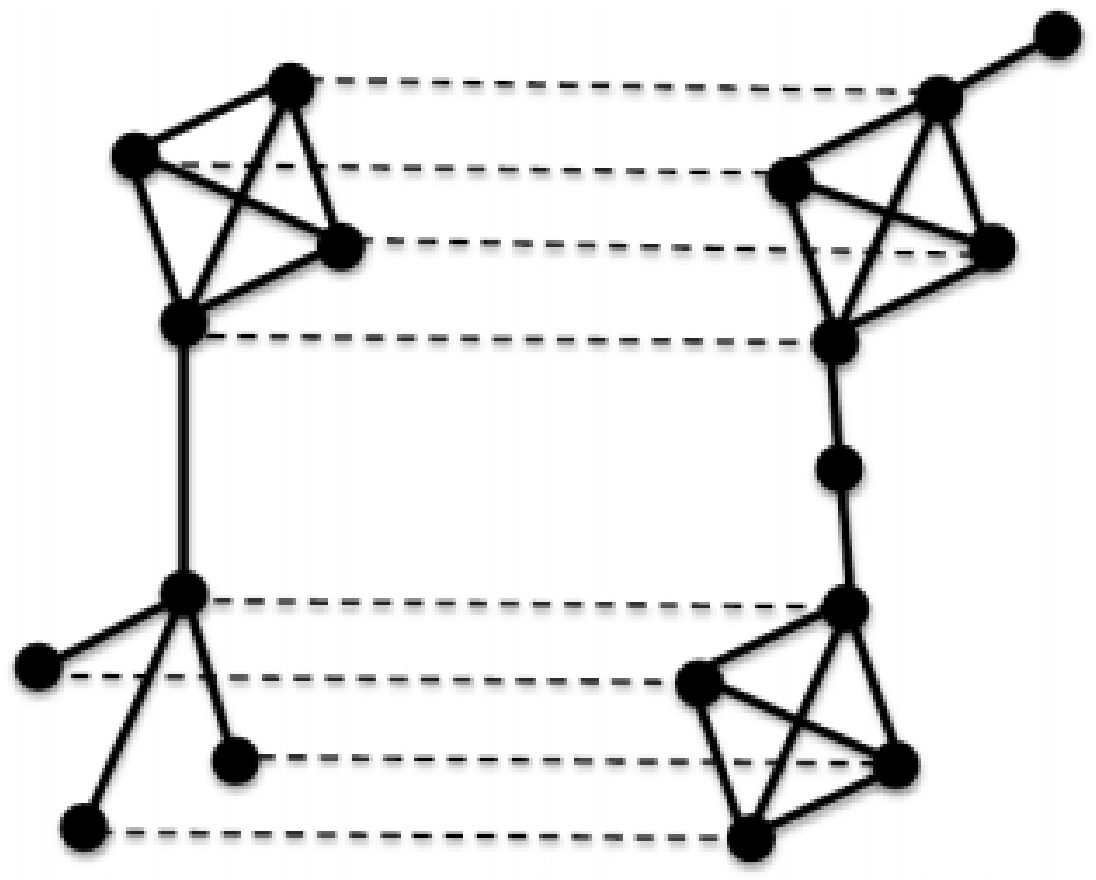}
\caption{Illustration of (a) LNA and (b) GNA, taken from
\cite{meng2015local}.}
\label{fig:lna_vs_gna}
\end{figure}

There exists two types of NA methods: local network alignment (LNA) and global network alignment (GNA). LNA aims to find a many-to-many node mapping (i.e., a node can be mapped
to one or more nodes from the other network) between networks of different species that identifies small but highly conserved subnetworks (Figure~\ref{fig:lna_vs_gna} (a)) \cite{Sharan2005networkblast,NETALIGNER,ALIGNNEMO,ALIGNMCL,LOCALALI}. On the other hand, GNA aims to find a one-to-one (injective) node mapping (i.e., every node in the smaller network is mapped to exactly one unique node in the larger network) that maximizes overall similarity of the compared networks, which often results in suboptimal conservation in local network regions (Figure~\ref{fig:lna_vs_gna} (b))\cite{IsoRank,MIGRAAL,GHOST,NETAL,Todor2013,ibragimov2013gedevo,GREAT,MAGNA,vijayan2015magna++,WAVE,LGRAAL,DualAligner,Clark09022015}.

We recently showed that LNA and GNA produce complementary results,
especially when gene or protein sequence information is included on
top of PPI network topological information during the alignment
construction process \cite{meng2015local}. This is because LNA and GNA
are designed to ``optimize'' different types of alignment quality: LNA
typically aims to ``optimize'' functional alignment quality, while GNA
aims to ``optimize'' topological alignment quality. It is very
challenging to design an NA (LNA or GNA) method that is of high
topological as well as functional quality, since it has been shown (by
others and us) that the topological versus functional fit between
aligned networks conflict to a larger extent than previously realized
\cite{Clark09022015, meng2015local, GHOST}.

Thus, we propose IGLOO, a new computational NA method that integrates
algorithmic components from both LNA and GNA in the hope of reconciling the
two NA types. IGLOO aims to ``inherit'' the advantages of both LNA and GNA,
i.e., the high functional quality of LNA and the high topological quality of
GNA. Specifically, given two networks and pairwise similarity scores
between their nodes (where the scores are computed via some node cost
function (NCF)), IGLOO takes a local alignment of high functional
quality generated by an existing LNA method (or a part of the local
alignment) as a seed alignment and then expands around the seed
alignment via an existing GNA method to increase topological alignment
quality. The difference between IGLOO and LNA is that IGLOO builds on
top of the given local alignment to improve its topological quality
(ideally without decreasing its functional quality). The difference
between IGLOO and GNA is that IGLOO uses as the seed a local alignment
of high functional quality (or a part of it) instead of using as the
seed just a single highly similar pair of nodes from the compared
networks (as GNA does), in order to improve functional quality of GNA
(ideally without decreasing its topological quality). As a result,
IGLOO's alignment is local in the sense that it allows for
many-to-many mapping between nodes of the two compared networks, just
as LNA does. Yet, its alignment is global in the sense that it allows
for mapping large conserved subgraphs across the compared networks,
just as GNA does.

In this paper, we comprehensively evaluate IGLOO against existing NA
methods both topologically and functionally. We study the following
state-of-the-art LNA methods: NetworkBLAST
\cite{Sharan2005networkblast}, NetAligner \cite{NETALIGNER}, AlignNemo
\cite{ALIGNNEMO}, and AlignMCL \cite{ALIGNMCL}. We study the following
state-of-the-art GNA methods: GHOST \cite{GHOST}, NETAL \cite{NETAL},
GEDEVO \cite{ibragimov2013gedevo}, MAGNA++ \cite{vijayan2015magna++},
WAVE \cite{WAVE}, and L-GRAAL \cite{LGRAAL}. We evaluate IGLOO against
these methods on four sets of PPI networks of varying interaction
types and confidence levels, and we do so with respect to proven
measures of topological and functional alignment quality (node
coverage combined with edge conservation, and precision and recall of
protein function prediction combined into F-score,
respectively)\cite{meng2015local}. We show that IGLOO produces a
better trade-off between topological and functional alignment quality
than the existing LNA and GNA methods. By this, we mean that across
all NA methods and network pairs, IGLOO is comparable or superior to
the existing methods both functionally and topologically in the
majority (62$\%$) of all cases.


\section{Methods}

\subsection{IGLOO algorithm}

IGLOO aims to ``inherit'' the advantages of both LNA and GNA, i.e., the
high functional quality of LNA and the high topological quality of
GNA. Intuitively, IGLOO achieves this by using the output of an LNA
method (i.e., its local alignment that is of high functional quality,
or a part of this alignment) as the seed within a GNA method that will
then expand the alignment around the seed (as GNA typically does) to
improve its topological quality. Typical GNA uses as the seed only a
single highly similar pair of nodes from the compared
networks. Instead, we vary the size of the seed from the entire local
alignment (i.e., 100\% of it) on one extreme (we expect this version
of IGLOO to resemble LNA the most) to only a single node pair (i.e.,
0\% of the local alignment) as the other extreme (we expect this
version of IGLOO to resemble GNA the most), with several
in-between-the-extremes versions of IGLOO that use as the seed a
certain portion (between 100\% and 0\%) of the local alignment (which
we expect will balance between high functional quality of LNA and high
biological quality of GNA).

More formally, the input to IGLOO is two networks: $G_1=(V_1, E_1)$ and $G_2=(V_2, E_2)$, where $V_i$ is the set of vertices in graph $i$, and $E_i$ the set of edges in graph $i$. The output of IGLOO is a list of aligned node pairs, where a node can theoretically appear in multiple aligned pairs. 
IGLOO transforms its input into output via the following algorithmic
steps: 
\begin{enumerate}
\vspace{-.08in}
\item Use a state-of-the-art LNA approach to find a local
alignment, i.e., a set of small conserved subnetworks of high
functional quality, which becomes the initial alignment. In this step,
all default parameters (including the NCF) of the existing LNA method
are used. 
\vspace{-.08in}
\item Compute a new NCF (as described below) that will be used in
the following steps to modify (decrease or increase) the current
alignment. 
\vspace{-.08in}
\item Iteratively decrease the size of the current alignment
by greedily removing (as explained below) the aligned node pair with
the lowest similarity score one at a time, until the user-specifi-
ed
alignment size is reached. This step is used to balance the
contributions of the LNA and the GNA during the alignment
process. Namely, the fewer node pairs are removed in step 3 from the
local alignment resulting from step 1 (i.e., the larger the seed size,
per our discussion above), the more similar IGLOO is to LNA (this
corresponds to the first extreme discussed above); the more node pairs
are removed (i.e., the smaller the seed size), the more similar IGLOO
is to GNA (this corresponds to the second extreme discussed
above). The resulting alignment becomes the new alignment. 
\vspace{-.08in}
\item Greedily expand (as explained below) around the current alignment by
iteratively adding node pairs with the highest similarity that have
remained unaligned up to this point, until no more node pairs can be
added to the alignment, meaning that each node in at least one of the
two compared networks has been aligned to some node(s). This step is
performed in order to find large conserved subnetworks of high
topological quality, just as GNA does. The resulting alignment becomes
the final alignment and the aligned node pairs (i.e., a many-to-many
node mapping) are returned as IGLOO's output. 
\end{enumerate}
\vspace{-\topsep} 
Next, we detail each
step.

\vspace{0.4cm}
\noindent\textbf{Step 1: Searching for small conserved subnetworks of high functional quality}
\vspace{0.2cm}

Since IGLOO must first generate a local alignment that has high
functional quality, IGLOO begins by using an LNA method. In order to
evaluate the robustness of IGLOO to the choice of LNA methods, we
evaluate our method on both AlignMCL and AlignNemo, two of the best
LNA methods \cite{meng2015local}. These methods rely on parameter
$\alpha$, which balances between the amount of topological versus
sequence information used in NCF during the alignment construction
process. For each method, we use the $\alpha$ value that results in
the highest functional quality of its local alignment, when varying
$\alpha$ from 0 (corresponding to using only sequence information to
compute NCF) to 1 (corresponding to using only topological information
to compute NCF) in increments of 0.1
\cite{meng2015local}. For IGLOO under AlignMCL, this results in
$\alpha=0.1$ for any network pair. For IGLOO under AlignNemo, this
results in $\alpha=0$ for any network pair except yeast-worm (Y2H$_1$)
and yeast-fly (Y2H$_1$), for which the best value of $\alpha$ is
$0.1$. Note that we describe the network data that we use in
Section~\ref{sec:data}.

\vspace{0.4cm}
\noindent\textbf{Step 2: Computing NCF to modify (decrease or increase) the current alignment in steps 3 and 4}
\vspace{0.2cm}

We compute NCF for steps 3 and 4 in the same way as NETAL \cite{NETAL} does, by combining node topological similarity (denoted as $TS$), node sequence similarity (denoted as $SS$), and interaction score (denoted as $IS$): $S=\beta (\alpha \cdot TS + (1-\alpha) \cdot SS) + (1-\beta) \cdot IS$. The first measure (i.e., $TS$) quantifies topological similarities between nodes from different networks using graphlet degree vector similarity (GDV-similarity) \cite{Milenkovic2008,meng2015local}. The second measure (i.e., $SS$) quantifies sequence similarities between nodes from different networks using normalized $E$-value \cite{meng2015local}. The third measure (i.e., $IS$) quantifies the similarity between two nodes as the number of edges that would be conserved if the two nodes were to be added next to the current alignment. The reason that we mimic NETAL's NCF is that compared to the existing GNA methods, NETAL results in the highest topological alignment quality, especially when using only topological information in NCF (corresponding to $\alpha=1$) \cite{meng2015local}. Note that NETAL's original implementation can use only topological information in NCF. For the purpose of IGLOO's development, we re-implement NETAL's NCF to also allow for using sequence similarity in NCF. Consequently, in terms of the $\alpha$ parameter, we tested both $\alpha=0$ and $\alpha=1$, and the results are similar. Hence, we simply choose $\alpha=1$ as was done in the original NETAL study. In terms of the $\beta$ parameter above, we use $\beta=0.001$ because IGLOO relies on NETAL's NCF and this value was suggested in the NETAL study \cite{NETAL}.

Unlike $SS$ and $TS$, which do not get updated throughout steps 3 and
4, $IS$ needs to be updated in each iteration of each of these steps
as IGLOO shrinks or expands the current alignment. When expanding the
current alignment, IGLOO mimics NETAL to update $IS$. When shrinking
the current alignment, IGLOO cannot mimic NETAL, since NETAL only
expands but never shrinks the current alignment. Thus, we first
discuss NETAL's strategy of updating $IS$ and then comment on how
IGLOO generalizes this strategy to update $IS$ when both expanding and
shrinking the current alignment.

For a node pair $i\in V_1$ and $j\in V_2$, NETAL computes its
interaction score, $IS(i, j)$, as follows. Under the assumption that
nodes $i$ and $j$ are to be aligned next and thus added to the current
alignment, NETAL computes: 1) $d_{ij}$, the number of conserved edges
that are incident to the node pair, 2) $p_i$, the expected value of
the number of \emph{candidate edges} (i.e., edges that are not
aligned) that are incident to $i$ and that will be conserved, and 3)
$p_j$, the expected value of the number of candidate edges that are
incident to $j$ and that will be conserved. The three values are
computed as follows. Let $N(x)$ be the neighbors of a node
$x$. Initially, $d_{ij}$ is set to zero since no conserved edges have
formed yet, and all edges that are incident to $i$ and $j$ are
candidate edges. Based on NETAL study, NETAL assumes that all
candidate edges that are incident to a node $x$ will be chosen to be
conserved with equal probability, and the probability that each
candidate edge $(x, x')$ where $x'\in N(x)$ will be conserved if $x$
is aligned to a random node is approximately $\frac{1}{|N(x')|}$. Let
$i'$ and $j'$ be two nodes from $N(i)$ and $N(j)$, respectively.
Since NETAL's assumption is that $i$ and $j$ are aligned, the
probability that edge $(i, i')$ will be conserved is
$\frac{1}{|N(i')|}$. Similarly, the probability that edge $(j, j')$
will be conserved is $\frac{1}{|N(j')|}$. Therefore, $p_i$ can be
measured by summing up the probabilities of all edges that are
incident to $i$ and that will be conserved: $p_i=\sum_{i'\in N(i)}
\frac{1}{|N(i')|}$. Similarly, $p_j=\sum_{j'\in N(j)}
\frac{1}{|N(j')|}$. After computing $d_{ij}$, $p_i$, and $p_j$,
$IS(i,j)$ is computed using Equation
\ref{equation:topologicalSimilarity}. Since $p_i$ and $p_j$ are not
greater than $|N(i)|$ and $|N(j)|$, respectively, $IS(i, j)$ is
normalized by the maximum node degree over all nodes from any of the
two compared networks (i.e., by $\max_{k\in V_1\cup V_2} \{|N(k)|\}$),
\begin{equation}
IS(i, j)=\frac{\min \{\sum_{i'\in N(i)} \frac{1}{|N(i')|}, \sum_{j'\in
N(j)} \frac{1}{|N(j')|}\}}{\max_{k\in V_1\cup V_2} \{|N(k)|\}}.
\label{equation:topologicalSimilarity}
\end{equation}
After $IS(i, j)$ is computed for the first time, NETAL updates its
value during each iteration of the alignment process from steps 3 and
4. Whenever two nodes $x\in V_1$ and $y\in V_2$ are aligned, NETAL
updates $IS(i,j)$ as follows: 1) if $x\in N(i)$ and $y\in N(j)$,
increase $d_{ij}$ by one; otherwise, do not update $d_{ij}$; 2) if $x$
is not aligned to any node from $V_2$ and $x\in N(i)$, decrease $p_i$
by $\frac{1}{|N(x)|}$; otherwise do not update $p_i$; 3) if $y$ is not
aligned to any node from $V_1$ and $y\in N(j)$, decrease $p_j$ by
$\frac{1}{|N(y)|}$; otherwise, do not update $p_j$; and 4) recompute
$IS(I, j)$ using Equation \ref{equation:interaction_score_update}. For
more details on how NETAL computes and updates $IS$, see \cite{NETAL}.
\begin{equation}
IS(i,j)=\frac{d_{ij} + \min \{ p_i, p_j\}}{\max_{k\in V_1\cup V_2} \{|N(k)|\}}
\label{equation:interaction_score_update}
\end{equation}

Now, we go back to explaining how IGLOO computes and updates
$IS$. When expanding the current alignment (steps 4) by adding a node
pair to it, IGLOO initially computes $IS$ based on the current
alignment using Equation \ref{equation:topologicalSimilarity} and
updates $IS$ just as NETAL does. When shrinking the current alignment
(step 3) by removing a node pair $x\in V_1$ and $y\in V_2$ from it,
IGLOO performs the following modifications: 1) if $x\in N(i)$ and
$y\in N(j)$, \emph{decrease} $d_{ij}$ by one; otherwise, do not update
$d_{ij}$; 2) if $x$ is aligned to $y$ only and $x\in N(i)$,
\emph{increase} $p_i$ by $\frac{1}{|N(x)|}$; otherwise do not update
$p_i$; 3) if $y$ is aligned to $x$ only and $y\in N(j)$,
\emph{increase} $p_j$ by $\frac{1}{|N(y)|}$; otherwise do not update
$p_j$; and 4) recompute $IS(i, j)$ using Equation
\ref{equation:interaction_score_update}.

\vspace{0.4cm}
\noindent\textbf{Step 3: Decreasing the size of the current alignment to balance between LNA and GNA}
\vspace{0.2cm}

IGLOO shrinks the current alignment greedily. Specifically, in each iteration, the node pair with the lowest similarity is removed from the current alignment, and the $IS$ is updated. The removal process terminates when the number of remaining aligned node pairs in the current alignment equals the user-specified alignment size $t$. We test five different values of $t$ to study its effect on alignment quality: 100\%, $75\%$, $50\%$, $25\%$, and $0\%$. These five values of $t$ will produce five alignments that IGLOO (i.e., its five versions) that will be expanded on in step 4. The five versions of IGLOO are: IGLOO 4, IGLOO 3, IGLOO 2, IGLOO 1 and IGLOO 0, respectively. IGLOO 4 takes the exact alignment produced from the given LNA method as the current alignment and expands around it, while IGLOO 0 does not use any of the local alignment and aligns the two networks from scratch, just as GNA does (Step 4). Therefore, we expect IGLOO 4 to be the most similar to LNA and IGLOO 0 the be the most similar to GNA, while the remaining versions of IGLOO will balance between high functional quality of LNA and high topological quality of GNA.

\vspace{0.4cm}
\noindent\textbf{Step 4: Searching for large conserved subnetworks of high topological quality}
\vspace{0.2cm}

IGLOO expands around the current alignment greedily to find large
conserved subnetworks of high topological quality, similar how GNA
works. In each iteration, IGLOO adds to the current alignment the node
pair from different networks that has remained unaligned up to that
point and that has the highest node similarity score, and then IGLOO
updates NCF scores accordingly. Each of the nodes that are aligned
cannot be used again in this expansion process. The expansion process
stops when no more node pairs can be added to the alignment. IGLOO
returns the latest current alignment as its final alignment. Note that
any expansion strategy (also called alignment strategy
\cite{MilenkovicACMBCB2013, Crawford2014}) can be used in IGLOO's step
4, including our recent alignment strategy called WAVE
\cite{WAVE}. We verified that using WAVE yields qualitatively
identical results as does using the above described expansion
strategy. Consequently, for brevity and simplicity, we leave out
discussion of WAVE's results and instead focus on results of the above
described strategy.

\subsection{Data}\label{sec:data}
We evaluate each NA (LNA, GNA, and IGLOO) method on four real-world
PPI network sets from our recent study \cite{meng2015local} containing
interactions of different types and confidence levels: 1) only yeast
two-hybrid physical PPIs, where each PPI is supported by at least one
publication (Y2H$_1$), 2) only yeast two-hybrid physical PPIs, where
each PPI is supported by at least two publications (Y2H$_2$), 3) all
physical PPIs, where each PPI is supported by at least one publication
(PHY$_1$), and 4) all physical PPIs, where each PPI is supported by at
least two publications (PHY$_2$). Each network set contains four PPI
networks of different species: \emph{S. cerevisiae} (yeast),
\emph{D. melanogaster} (fly), \emph{C. elegans} (worm), and
\emph{H. sapiens} (human). For each network, we use its largest
connected component, just as in \cite{meng2015local}. We do not
include those pairs involving Y2H$_2$ and PHY$_2$ networks of worm and
yeast, since these four networks are extremely small and sparse, with
random-like topology.

AlignNemo is able to produce an alignment for six of the
aforementioned network pairs (it cannot successfully run for the other
network pairs, for reasons discussed in
\cite{meng2015local}). Thus, since IGLOO is partly based on AlignNemo,
in order to fairly evaluate IGLOO against the existing NA methods, we
focus on the six network pairs that AlignNemo is able to run on. The
network pairs are: yeast-fly (Y2H$_1$), yeast-worm (Y2H$_1$), worm-fly
(Y2H$_1$), yeast-human (Y2H$_2$), yeast-worm (PHY$_1$), and fly-worm
(PHY$_1$). The size of each network is shown in Table~\ref{tab:networks}. For more
details on each data set, see
\cite{meng2015local}.

\begin{table}[h!]
\centering
\caption{The number of nodes and edges for each network used in this study.}

\begin{tabular}{|l|l|l|}
\hline
\textbf{Network}         & \textbf{\# of nodes} & \textbf{\# of edges} \\ \hline
\textbf{yeast (Y2H$_1$)} & 3,427          & 11,348         \\ \hline
\textbf{fly (Y2H$_1$)}   & 7,094          & 23,356         \\ \hline
\textbf{worm (Y2H$_1$)}  & 2,871          & 5,194          \\ \hline
\textbf{yeast (Y2H$_2$)} & 744            & 966            \\ \hline
\textbf{human (Y2H$_2$)} & 1,191          & 1,567          \\ \hline
\textbf{yeast (PHY$_1$)} & 6,168          & 82,368         \\ \hline
\textbf{fly (PHY$_1$)}   & 7,885          & 36,271         \\ \hline
\textbf{worm (PHY$_1$)}  & 3,003          & 5,501          \\ \hline
\end{tabular}

\label{tab:networks}
\end{table}

\subsection{Alignment quality measures}
We evaluate each NA (LNA, GNA, and IGLOO) method in terms of both
topological and functional alignment quality. We focus on node
coverage (NCV) combined with the generalized symmetric substructure
score (GS$^3$) measure of edge conservation as a measure of
topological alignment quality, where the combined topological measure
is denoted as NCV-GS$^3$. Also, we focus on precision (P-FP) and
recall (R-PF) of protein function prediction combined into F-score as
a measure of functional alignment quality, where the combined
functional measure is denoted as F-PF. We use these measures because
they are already proven evaluation criteria for both LNA and GNA that
can compare the two fairly \cite{meng2015local}.

Intuitively, NCV-GS$^3$ quantifies the size of the given alignment in terms of the amount of both conserved nodes (NCV) and conserved edges (GS$^3$). Let $f$ be an alignment between two graphs $G_1(V_1,E_1)$ and $G_2(V_2,E_2)$, and let $G'_1(V'_1,E'_1)$ and $G'_2(V'_2,E'_2)$ be subgraphs of $G_1$ and $G_2$ that are induced on node sets $f(V_2)$ and $f(V_1)$. NCV is the percentage of nodes from $G_1$ and $G_2$ that are also in $G'_1$ and $G'_2$ (i.e., $\frac{|V'_1|+|V'_2|}{|V_1|+|V_2|}$ ). GS$^3$ is the percentage of conserved edges out of the total of both conserved and non-conserved edges. NCV-GS$^3$ is the geometric mean of NCV and GS$^3$. 

Before we define F-PF, we note that this measure is computed with
respect to Gene Ontology (GO) gene-function annotation data
\cite{MIGRAAL}. We only use gene-GO term annotations that
have been obtained experimentally. That is, we discard those
functional annotations that have been obtained e.g., computationally
via sequence alignment. We do this because the NA methods that we
evaluate already use sequence information within NCF when producing
their alignments, and thus evaluating such alignments with respect to
sequence-based functional annotations would lead to a circular
argument, which is undesirable
\cite{MIGRAAL}.

Now, we go back to defining F-PF. This measure quantifies how similar
the aligned nodes are in terms of their functions
\cite{meng2015local}. To compute F-PF, we first hide proteins' true GO
terms and then predict the proteins' GO terms based on GO terms of
their aligned counterpart(s) \cite{meng2015local}. Next we compute the
P-PF and R-PF of the resulting predicted GO terms with respect to the
true GO terms. Finally, F-PF is the harmonic mean of P-PF and R-PF.

For more details on the NCV-GS$^3$ and F-PF measures, see \cite{meng2015local}.

\section{Results and discussion}
We evaluate IGLOO against the existing NA methods considered in our study in terms of both alignment quality (Section \ref{sec:quality_analysis}) and running time (Section \ref{sec:running_time_analysis}).

\subsection{Alignment quality method comparison}
\label{sec:quality_analysis}
Here, we show results for IGLOO when considering the total of 10 IGLOO
versions: IGLOO 0--4 when each of AlignMCL and AlignNemo are used in
step 1 of the IGLOO algorithm. When we compare the different methods
(the existing LNA and GNA methods, and the 10 IGLOO versions), for a
given network pair and a given existing NA method, we obtain results
for four possible cases: 1) IGLOO is comparable or superior
both topologically and functionally, meaning that at least one version
of IGLOO is comparable or superior both topologically and
functionally; 2) IGLOO is comparable or superior only
functionally but not topologically, meaning that no version of IGLOO
is comparable or superior both topologically and functionally, and at
least one version of IGLOO is comparable or superior only functionally
but not topologically; 3) IGLOO is comparable or superior
only topologically but not functionally, meaning that none of the
versions of IGLOO are comparable or superior both topologically and
functionally, and at least one version of IGLOO is comparable or
superior only topologically but not functionally; and 4)
IGLOO is inferior both topologically and functionally, meaning that
all versions of IGLOO are inferior. Note that cases 2 and 3 could
occur at the same time, since it is possible that some version of
IGLOO is comparable or superior only topologically but not
functionally, while another version is comparable or superior only
functionally but not topologically. However, none of cases 1, 2, and
4, or cases 1, 3, and 4, can occur at the same time.

Regarding the NCFs of the existing NA methods considered in our study,
different $\alpha$ values (i.e., where $\alpha$ balances between the
amount of topological versus sequence information in NCF) might result
in different alignment quality. Therefore, for each existing NA method
and each network pair, we choose the $\alpha$ value that results in
the best trade-off between topological and functional quality when
varying $\alpha$ from 0 to 1 in increments of 0.1
\cite{meng2015local}. We measure the trade-off between the
two quality types by computing their geometric mean. We report results
only for the $\alpha$ value that results in the maximum geometric
mean.

\begin{figure}[h!]
\centering
\includegraphics[width=.77\columnwidth]{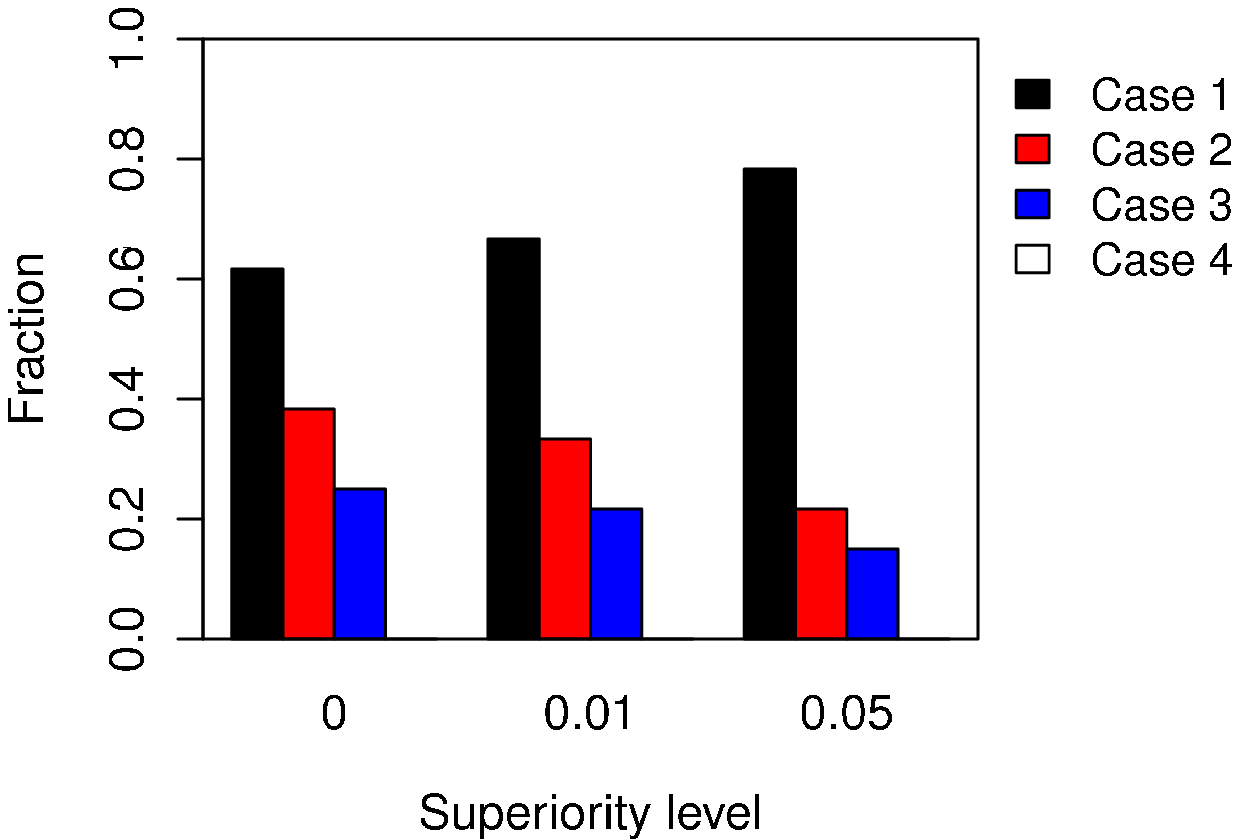}\\
\textbf{(a)} LNA and GNA combined\\
\vspace{0.2cm}
\includegraphics[width=0.77\columnwidth]{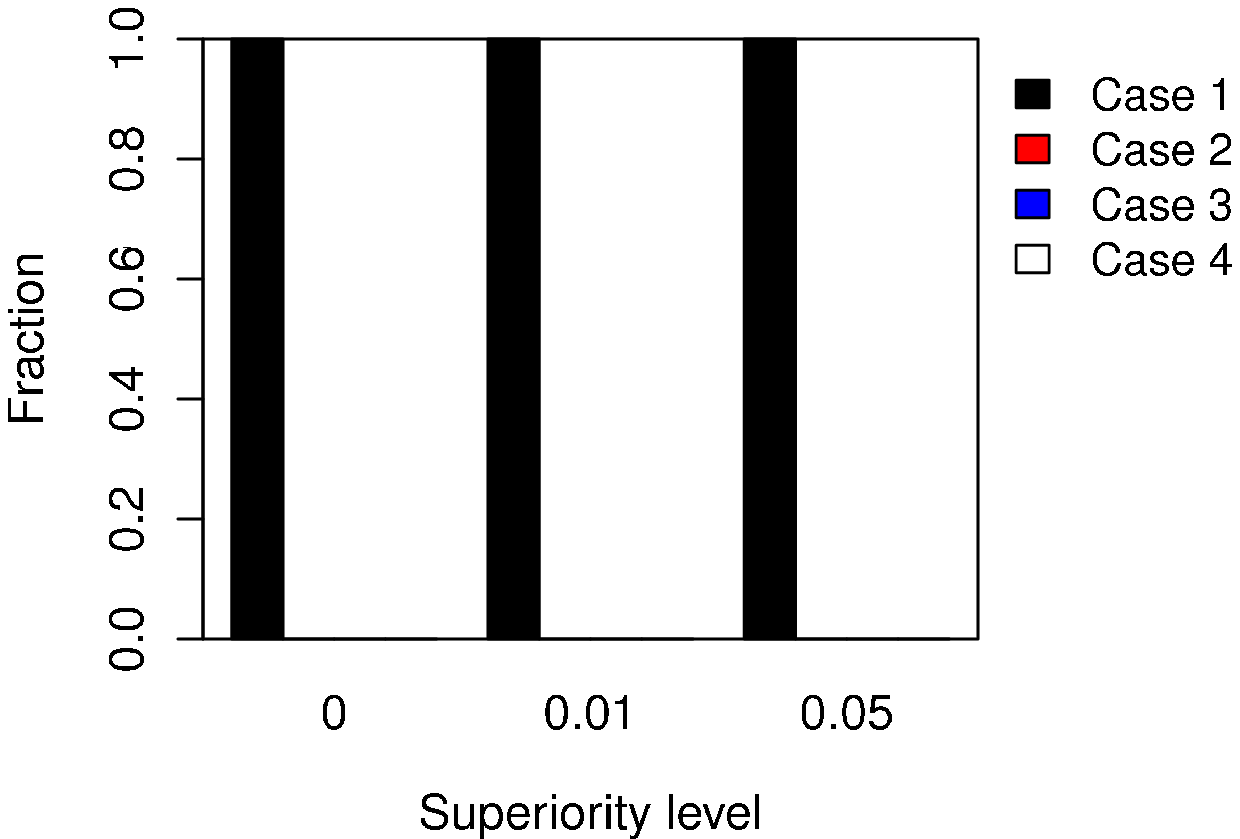}\\
\textbf{(b)} LNA \\
\vspace{0.2cm}
\includegraphics[width=0.77\columnwidth]{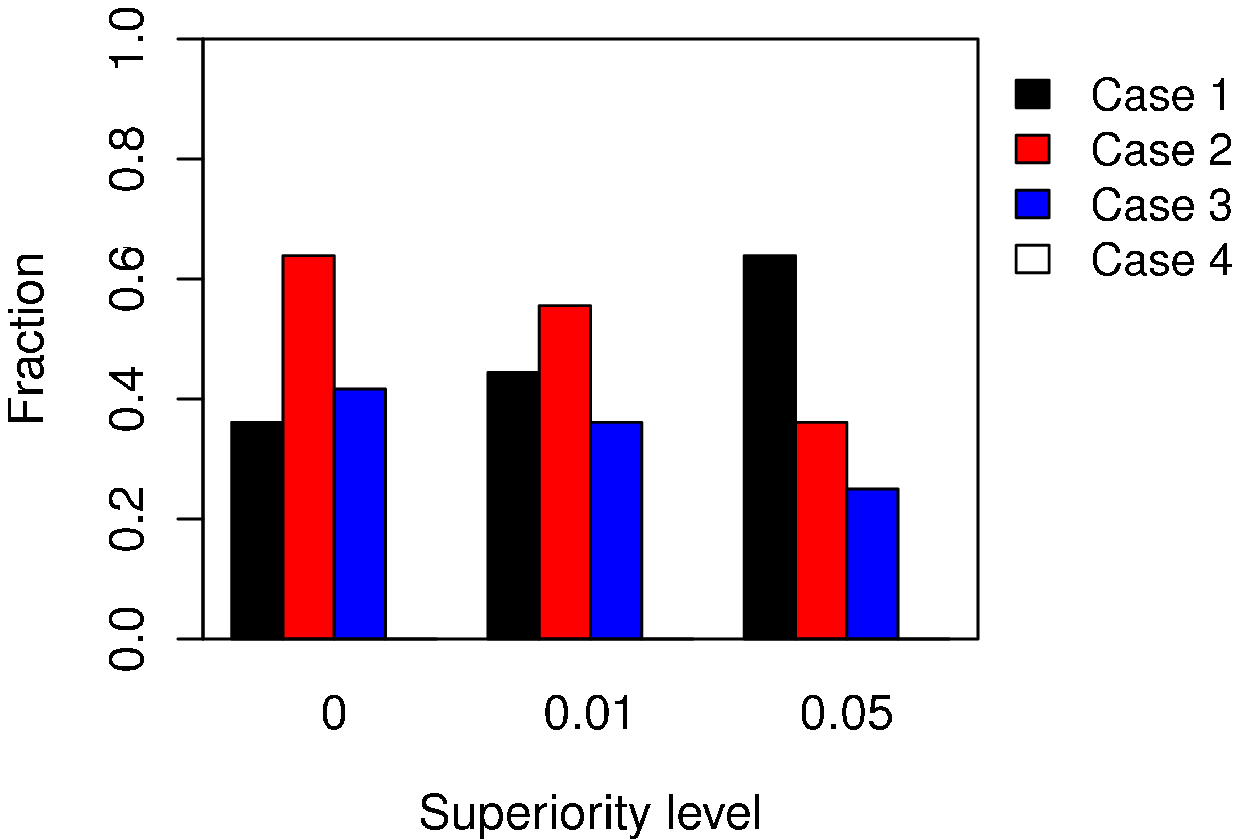}\\
\textbf{(c)} GNA 
\caption{Overall comparison of IGLOO (the best of its versions) and 
\textbf{(a)} LNA and GNA combined, \textbf{(b)} LNA, and \textbf{(c)} GNA, 
when considering 10 different IGLOO versions: IGLOO 0-4 for each of
AlignMCL and AlignNemo used in step 1 of the algorithm. The
comparison is shown for three different method ``superiority levels''
(denoted as $p$): 0\%, 1\%, and 5\%. By a ``superiority level'', we
mean the following. Given two methods $A$ and $B$ with alignment
quality scores $x$ and $y$, respectively, if $\frac{|x-y|}{max(x, y)}
\leq p$, we say that $A$ and $B$ are comparable; otherwise, if $x$ is
greater/less than $y$, we say that $A$ is superior/inferior to
$B$. For a given network pair and a given existing method, only the
best version of IGLOO is considered. The four cases are: 1) IGLOO is
comparable or superior both topologically and functionally; 2) IGLOO
is comparable or superior only functionally but not topologically; 3)
IGLOO is comparable or superior only topologically but not
functionally; and 4) IGLOO is inferior both topologically and
functionally. The \emph{y}-axes indicate the percentage of the
combinations of the existing NA methods and networks pairs for which
the given case occurs.}
\label{fig:comparison_combined}
\end{figure}

\begin{figure}[ht!]
\centering
\begin{minipage}{0.38\textwidth}
\centering
\textbf{(a)}\includegraphics[width=\textwidth]{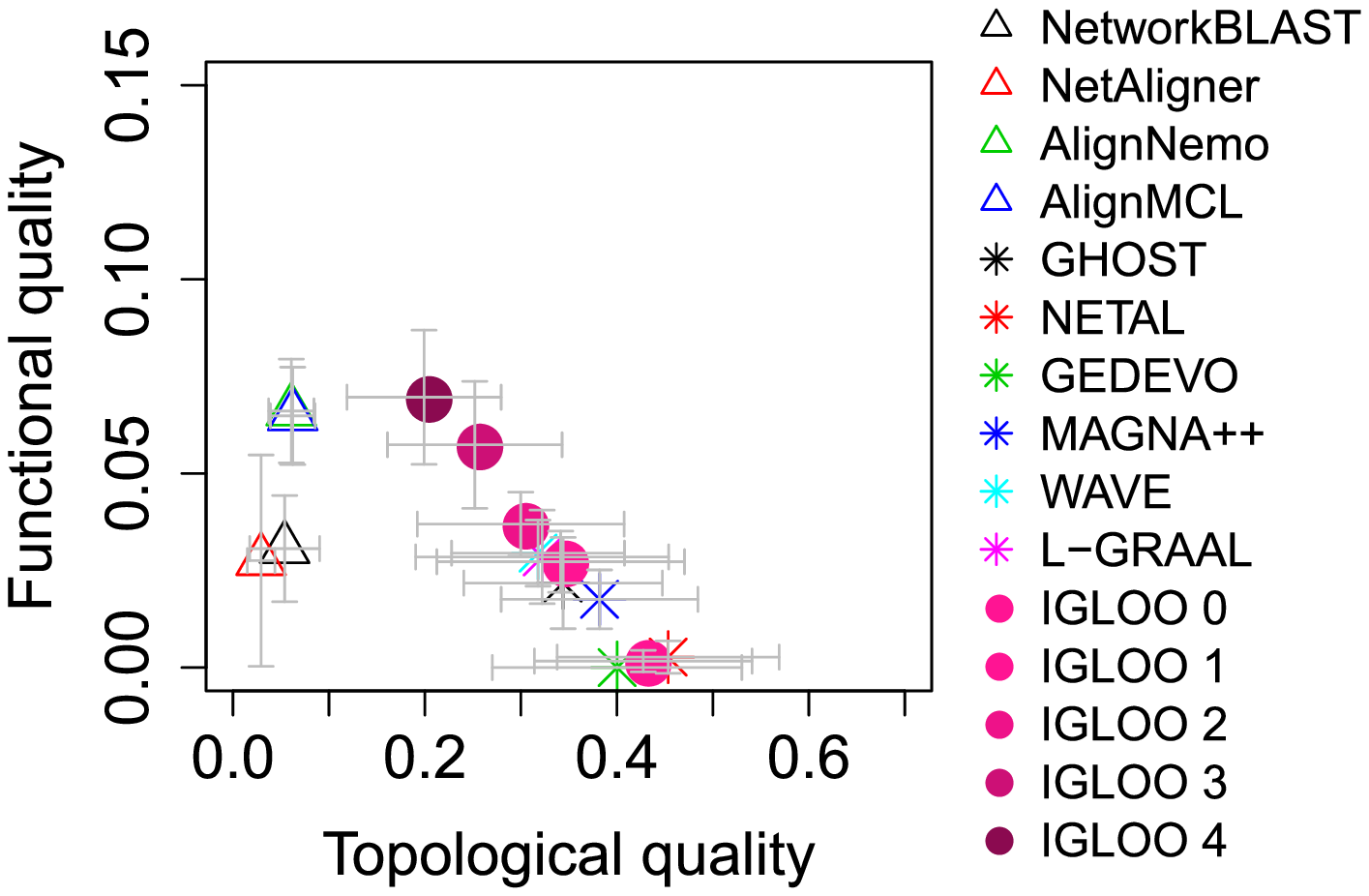}
\end{minipage}\\
\begin{minipage}{0.38\textwidth}
\centering
\textbf{(b)}\includegraphics[width=.92\textwidth]{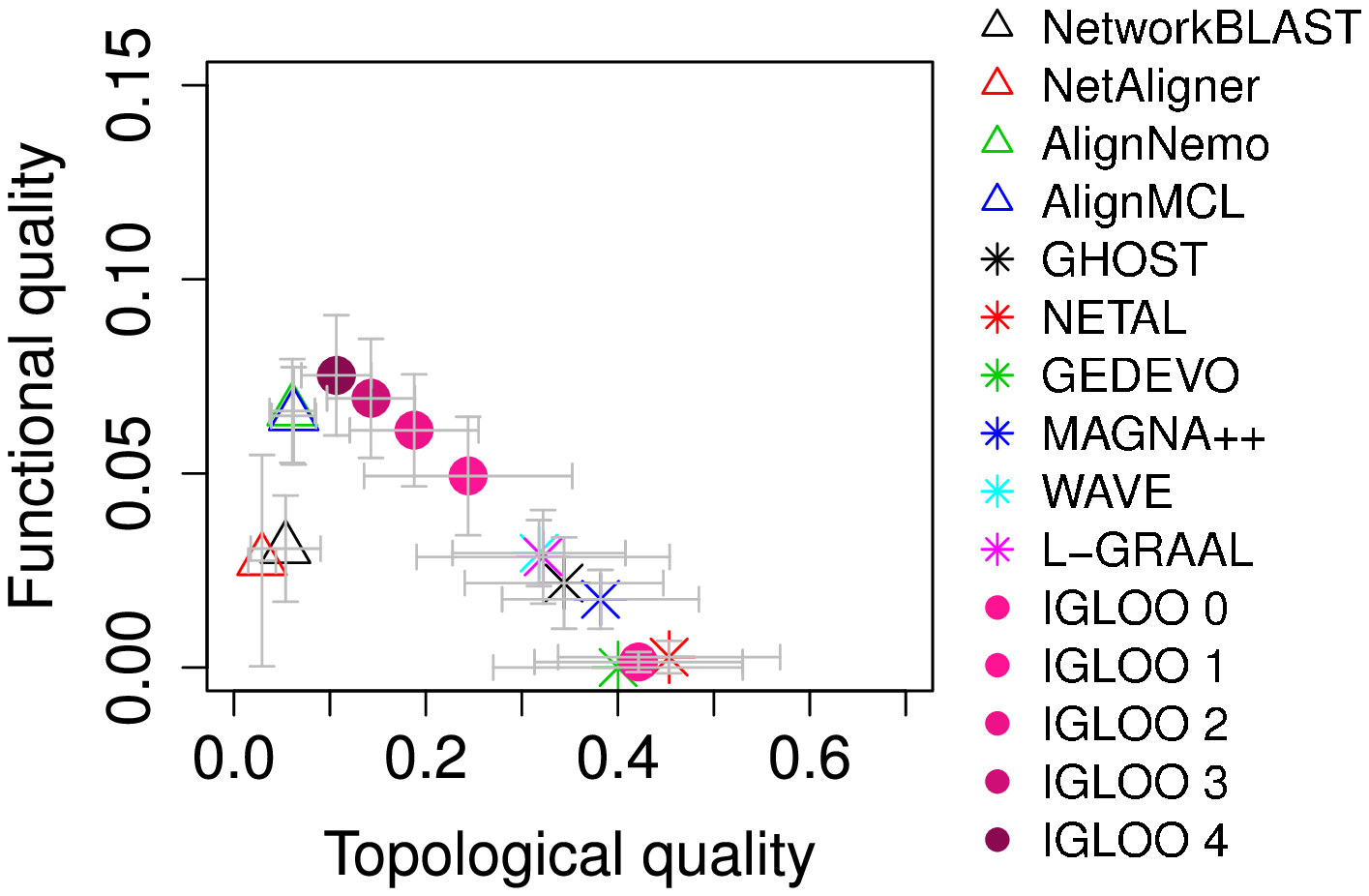}
\end{minipage}
\caption[Topological (NCV-GS$^3$; $x$-axis) and functional (F-PF; $y$-axis) alignment quality for each NA method (IGLOO, LNA, and GNA) across all compare network pairs]{Topological (NCV-GS$^3$; $x$-axis) and functional (F-PF; $y$-axis) alignment quality for the existing LNA methods (triangles), existing GNA methods (stars), and IGLOO versions (circles), averaged over all aligned network pairs, when considering \textbf{(a)} AlignMCL and \textbf{(b)} AlignNemo in the first step of the IGLOO algorithm. For detailed results for each network pair individually, see Supplementary Figures \ref{fig:top_bio_alignmcl1} and \ref{fig:top_bio_alignnemo}.}
\label{fig:top_bio_average}
\end{figure}

Our findings are as follows. Overall, IGLOO is comparable or superior
to the existing NA methods considered in our study (Figures
\ref{fig:comparison_combined} (a) and
\ref{fig:top_bio_average}). Specifically, when considering all
combinations of the existing NA methods and networks pairs, in 62\% of
them, IGLOO is comparable or superior both topologically and
functionally (case 1). In 38\% of the combinations, IGLOO is
comparable or superior only functionally but not topologically (case
2). In 25\% of the combinations, IGLOO is comparable or superior only
topologically but not functionally (case 3). IGLOO is never inferior
both topologically and functionally (case 4). Note that 25\% of the
combinations are in the overlap of cases 1 and 2. When considering two
given methods to be comparable if their alignment quality scores are
within 1\% or 5\% of each other, IGLOO is even more comparable or
superior both topologically and functionally, in up to 78\% of all
cases (Figure \ref{fig:comparison_combined} (a)). That is, often, when
the existing methods are comparable or superior to IGLOO, their
superiority is only within 1\% or 5\% of IGLOO alignment
quality. Equivalent results when considering only the five
AlignMCL-based IGLOO versions and only the five AlignNemo-based IGLOO
versions (as opposed to all 10 versions of IGLOO) are shown in
Supplementary Figures \ref{fig:comparison_alignmcl} and
\ref{fig:comparison_alignnemo}, respectively.

\begin{figure}[h!]
\centering
\includegraphics[width=0.5\columnwidth]{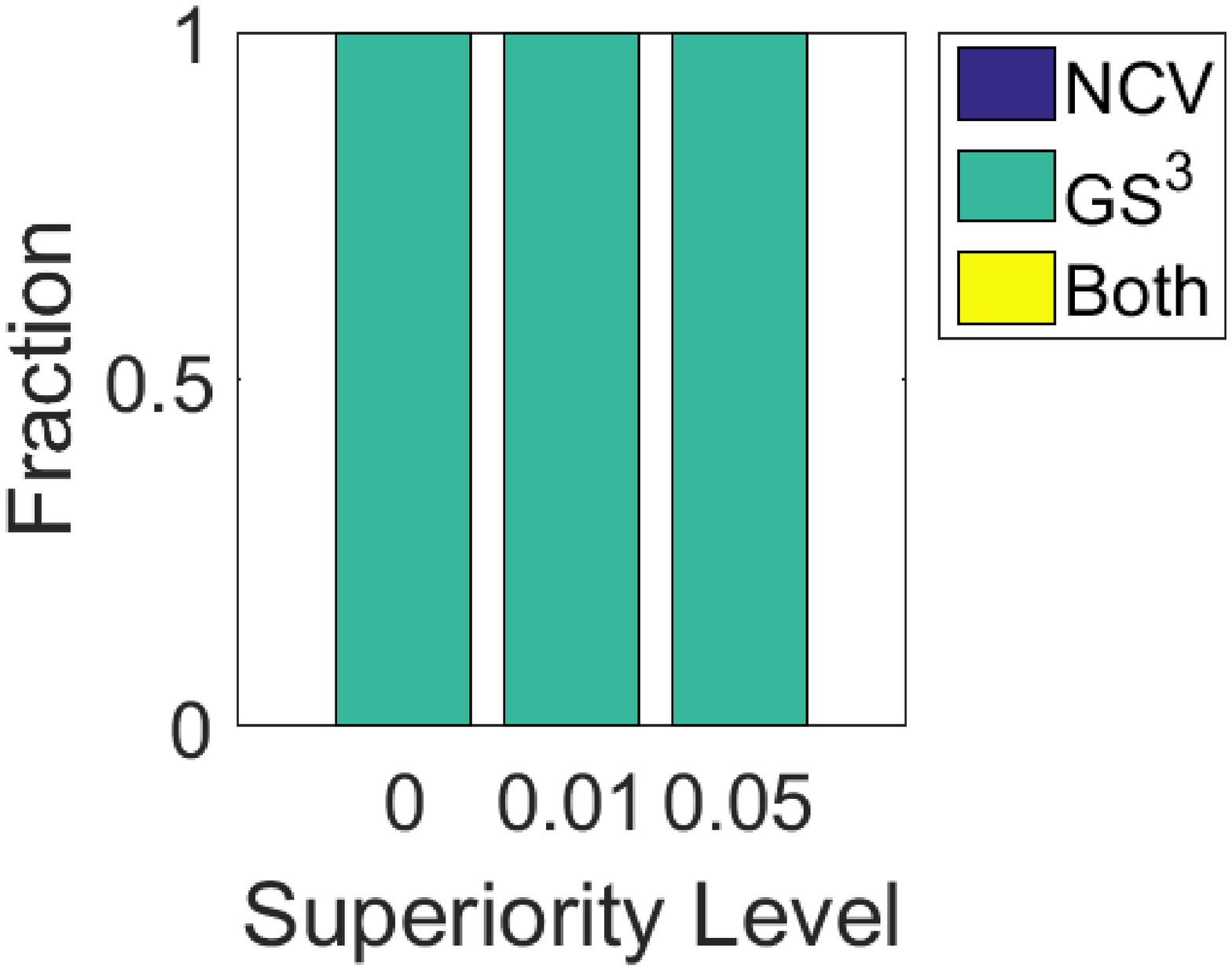}
\hspace{-0.2cm}
\includegraphics[width=0.5\columnwidth]{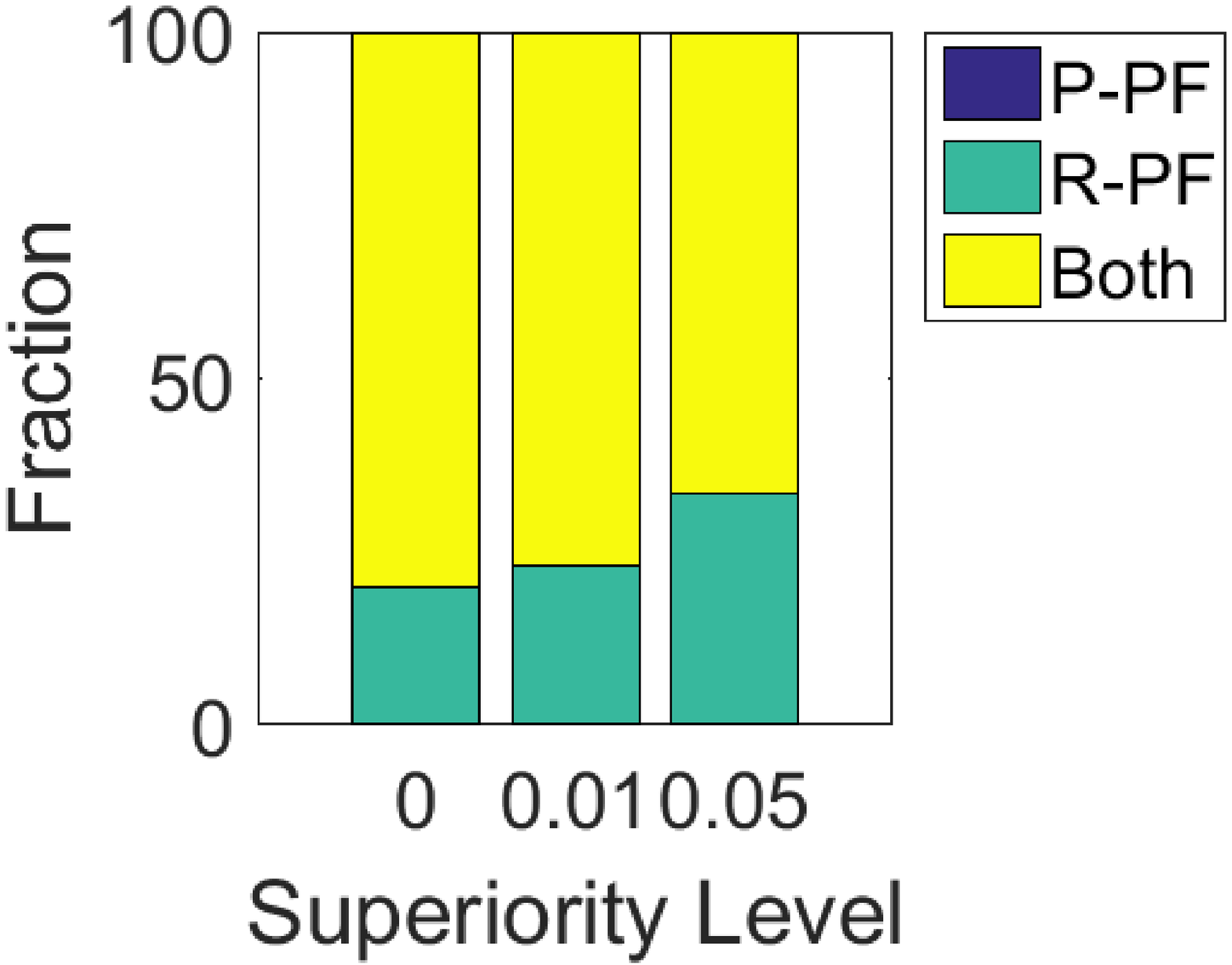}\\
\hspace{-0.4cm}\textbf{(a)}\hspace{3.9cm}\textbf{(b)}
\caption{Reasons behind the superiority of the existing methods 
over IGLOO in (a) case 2 and (b) case 3. For case 2 (whenever the
existing methods outperform IGLOO topologically but not functionally),
we show the percentage of all instances in which the topological
superiority of the existing methods is with respect to NCV only,
GS$^3$ only, or both. Similarly, for case 3 (whenever the existing
methods outperform IGLOO functionally but not topologically), we show
the percentage of all instances in which the functional superiority of
the existing methods is with respect to P-PF only, R-PF only, or
both. }
\label{fig:detailed_scores}
\end{figure}

Importantly, for case 2, whenever IGLOO is comparable or superior to
the existing methods functionally but not topologically, or in other
words whenever the existing methods outperform IGLOO topologically but
not functionally, the topological superiority of the existing methods
(in terms of NCV-GS$^3$) comes only from GS$^3$ but not NCV
(Figure~\ref{fig:detailed_scores} (a)). Similarly, for case 3,
whenever IGLOO is comparable or superior to the existing methods
topologically but not functionally, or in other words whenever the
existing methods outperform IGLOO functionally but not topologically,
the functional superiority of the existing methods (in terms of F-PF)
comes only from recall (R-PF) but not precision (P-PF) in 20-33.4$\%$
of all cases (Figure~\ref{fig:detailed_scores} (b)); for biological
scientists, precision of protein function prediction (making as
accurate predictions as possible, even if few of them) is likely more
important than R-PF (making as many predictions as possible, even if
less accurate).

Next, we zoom into these results to compare IGLOO to each of LNA and
GNA individually (Figure \ref{fig:comparison_combined} (b) and (c),
respectively).

The comparison results against LNA are as follows. IGLOO is comparable
or superior to all of the existing LNA methods considered in our study
both topologically and functionally for all network pairs. When
measuring the within 1\% or within 5\% accuracy (as described above),
IGLOO remains comparable or superior both topologically and
functionally. That is, IGLOO is at least 5\% better than any of the
existing LNA methods, both functionally and topologically. Thus, since
IGLOO improves both topological and functional alignment quality of
the existing LNA methods, at the minimum, IGLOO's contribution is the
new best LNA method.

The comparison results against GNA are as follows. When considering all combinations of the existing GNA methods
and networks pairs, in 36\% of them, IGLOO is comparable or superior
both topologically and functionally (case 1). In 64\% of the
combinations, IGLOO is comparable or superior only functionally but
not topologically (case 2). In 42\% of the combinations, IGLOO is
comparable or superior only topologically but not functionally (case
3). IGLOO is never inferior both topologically and functionally (case
4). Note that 42\% of the combinations are in the overlap of cases 1
and 2. When measuring the within 1\% or within 5\% accuracy, similar
trends hold, except that now IGLOO is comparable or superior in up to
64\% of all cases both topologically and functionally. That is, often,
when the existing GNA methods are comparable or superior to IGLOO,
their superiority is only within 1\% or 5\% of IGLOO's alignment
quality. Further, over all combinations of the existing GNA methods
and networks pairs in which IGLOO improves functional quality of the
GNA methods but lowers their topological quality (case 2), the average
improvement in functional quality is 331\% (standard deviation of
276\%), while the average decrease in topological alignment quality is
only 40\% (standard deviation of 11\%). Thus, IGLOO gains more than it
loses. Over all combinations of the existing GNA methods and networks
pairs in which IGLOO improves topological quality of the GNA methods
but lowers their functional quality (case 3), the average improvement
in topological quality is 18\% (standard deviation of 12\%), while the
average decrease in functional alignment quality is 73\% (standard
deviation of 31\%). Therefore, IGLOO overall beats the existing GNA
methods for case 2, while the existing GNA methods beat IGLOO for case
3.

Equivalent results when considering only the five AlignMCL-based IGLOO
versions and only the five AlignNemo-based IGLOO versions (as opposed
to all 10 versions of IGLOO) are shown in Supplementary Figures
\ref{fig:comparison_alignmcl} and \ref{fig:comparison_alignnemo}, respectively.

We conclude our discussion on comparison of the different methods in
terms of their alignment accuracy by summarizing the performance of
each method over all analyzed network pairs (Figure
\ref{fig:top_bio_average}) and in particular by computing the
statistical significance of the improvement of one method over another
(where we use the paired $t$-test to compare alignment scores of two
methods of interest over all network pairs). Based on these results,
we comment on which version of IGLOO (out of IGLOO 0-4) is the best.

For LNA, there is a version of IGLOO, in particular IGLOO 4 under
AlignNemo, which is superior in a statistically significantly manner ($p$-value
$<$ 0.05) to each considered LNA method in terms of both topological
and functional alignment quality (Figure
\ref{fig:top_bio_average}). In addition, IGLOO 3-4 under
AlignMCL and IGLOO 2-4 under AlignNemo are statistically significantly
superior to two of the four considered LNA methods (NetAligner and
NetworkBlast).

For GNA, no version of IGLOO is superior in a statistically significantly manner
to any existing GNA method in terms of both topological and functional
alignment quality. However, we note that: 1) IGLOO is still superior
to the existing GNA methods in many cases, as shown in Figure
\ref{fig:comparison_combined}, it is just that its superiority is 
not statistically significant, and 2) importantly, none of the
existing GNA methods is statistically significantly superior to any
version of IGLOO in terms of both topological and functional alignment
quality. Clearly, each of IGLOO and an existing GNA method is at best
statistically significantly superior either functionally or
topologically, but not both. So, for GNA, we split the discussion into
two cases: 1) when IGLOO is statistically significantly better than
the given existing GNA method in terms of only functional alignment
quality, and 2) when IGLOO is statistically significantly better than
the given existing GNA method in terms of only topological alignment
quality.

For the first case above (IGLOO is statistically significantly
superior to GNA only functionally), IGLOO 3-4 under any of AlignMCL or
AlignNemo are statistically significantly superior to each considered
GNA method. In addition, IGLOO 1-2 under AlignNemo are statistically
significantly superior to each considered GNA method except
L-GRAAL. The remaining versions out of IGLOO 1-4 under either AlignMCL
or AlignNemo are statistically significantly superior to at least one
of the considered GNA methods. Of all IGLOO versions, only IGLOO 0
under any of AlignMCL and AlignNemo is never statistically
significantly superior to any of the existing GNA methods in terms of
functional alignment quality.

For the second case above (IGLOO is statistically significantly
superior to GNA only topologically), no version of IGLOO beats NETAL
or MAGNA++. This is not surprising, because these are among the best
GNA methods in terms of topological alignment quality
\cite{meng2015local}. For the remaining four GNA methods, IGLOO 0 under AlignMCL is statistically significantly superior to each of the four
methods. Also, IGLOO 0 under AlignNemo is statistically significantly
superior to each of the four methods except GEDEVO. Finally, IGLOO 1
under AlignNemo is statistically significantly superior to L-GRAAL. No
other version of IGLOO is statistically significantly superior to any
GNA method topologically.

In summary, in terms of functional alignment quality, IGLOO 4 is the
strongest compared to both LNA and GNA, and it is followed by IGLOO 3
and IGLOO 2. In terms of topological alignment quality, IGLOO 4 is the
strongest compared to LNA, and it is followed by IGLOO 3 and IGLOO 2,
while IGLOO 0 is the strongest compared to GNA, and it is followed by
IGLOO 1.

\subsection{Running time method comparison}

\label{sec:running_time_analysis}
Here, we compare IGLOO when using AlignMCL in the first step of the
algorithm against each of the existing LNA and GNA methods in terms of
computational complexity. We run all NA methods on the same Linux
machine with 64 CPU cores (AMD Opteron (tm) Processor 6378) and 512 GB
of RAM. All methods can run on a single core with the exception of
GHOST, which can run on at least two cores. Three of the existing GNA
methods (GHOST, GEDEVO, and MAGNA++) can run on multiple cores. The
maximum number of cores that the parallelizable methods can use is
bounded by the number of cores that our machine has. We analyze the
methods' entire running times, which encompass both computing node
similarities and constructing alignments. Also, we measure only
running times needed to construct alignments, ignoring the time needed
to precompute node similarities. We show the results for worm and
yeast PPI networks of Y2H$_1$ type, since both networks are relatively
small, and even the slowest NA method could finish aligning the two
networks on a single core within a reasonable time (within one
day). For any other network pair, it could take much longer time for
the slowest method to finish.

\begin{figure}[h!]
\centering
\textbf{(a)}\includegraphics[width=0.44\textwidth]{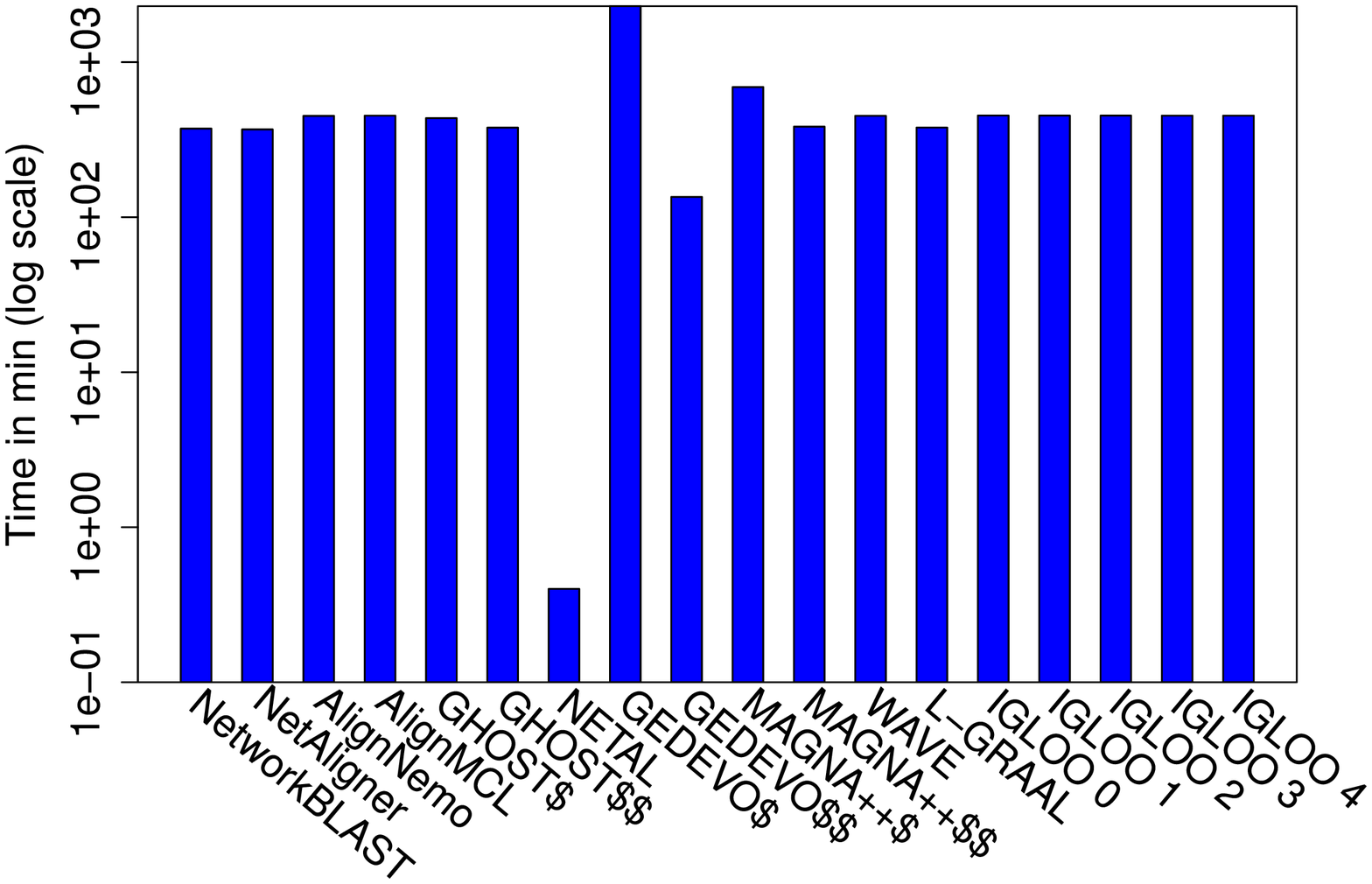}
\textbf{(b)}\includegraphics[width=0.44\textwidth]{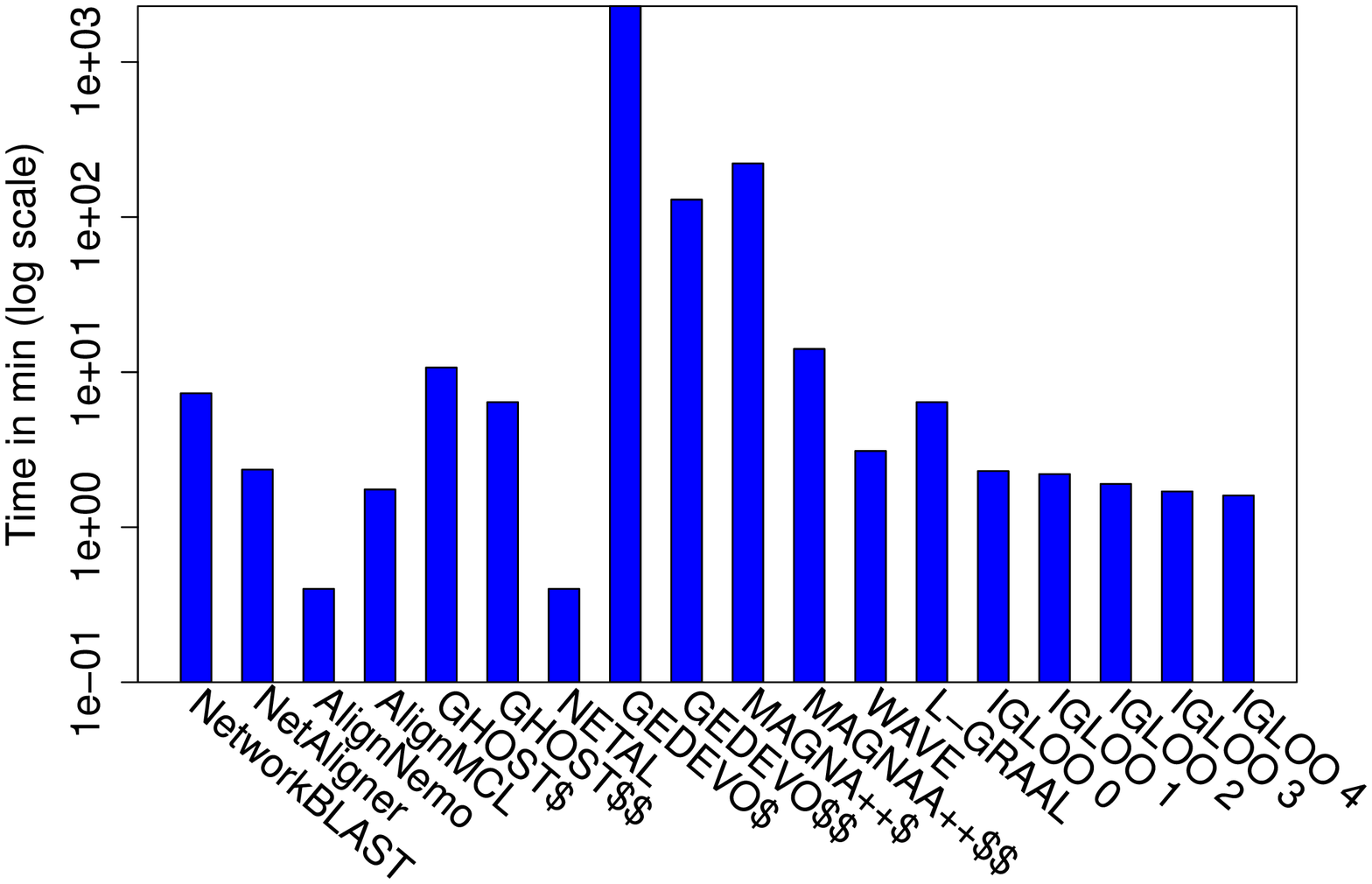}
\caption{Representative running time comparison of the different NA
methods, for \textbf{(a)} the entire running times and \textbf{(b)}
only the times for constructing alignments. For each method that is
parallelizable, its single-core version is marked with a `\$' symbol,
and its 64-core version is marked with a `\$\$' symbol. All other
methods are run on a single core. Results are shown for using AlignMCL
in the first step of IGLOO algorithm.}
\label{fig:running_time}
\end{figure}

Regarding the entire running times, the findings are as follows
(Figure \ref{fig:running_time} (a)). Since IGLOO uses AlignMCL and
NETAL within its algorithm, it is not surprising that IGLOO is
(slightly) slower than these two methods. Of the remaining methods,
IGLOO is faster than two methods (i.e., serial GEDEVO and serial
MAGNA++), it is relatively comparable to three methods (i.e.,
AlignNemo, serial GHOST, and WAVE), and it is slower than six methods
(i.e., NetworkBLAST, NetAligner, parallelized GHOST, parallelized
GE-
DEVO, parallelized MAGNA++, and L-GRAAL).

Regarding only the times for computing alignments, the findings are as
follows (Figure \ref{fig:running_time} (b)). Again, IGLOO is
(slightly) slower than AlignMCL and NETAL. Of the remaining methods,
IGLOO is faster than eight methods (NetworkBLAST, serial GHOST,
parallelized GHOST, serial GEDEVO, parallelized GEDEVO, serial
MAGNA++, parallelized MAGNA++, and L-GRAAL), it is relatively
comparable to two methods (NetAligner and WAVE), and it is slower than
one method (AlignNemo). Importantly, all methods except perhaps serial
and parallelized GEDEVO and serial MAGN-
A++ have reasonably low running
times.

\vfill

\section{Conclusion}
We propose a new NA method, IGLOO, which aims to combine the
advantages of both LNA and GNA in order better balance between
functional and topological alignment quality. We demonstrate that
IGLOO outperforms all considered LNA methods with respect to both
alignment quality types, it also outperforms the considered GNA
methods in many cases.

IGLOO is generalizable as it can include any existing LNA and GNA
methods into its algorithm. (The existing methods we test are simply a
proof of concept of combining LNA with GNA.) As the field of NA
evolves, including newer and more sophisticated methods could further
improve the alignment quality of IGLOO.

\section{Acknowledgements}
\noindent
This material is based upon work supported by the National Science Foundation under Grant No. CAREER CCF-1452795 and
CCF-1319469.

\newcommand{\beginsupplement}{%
\setcounter{table}{0}
\renewcommand{\thetable}{S\arabic{table}}%
\setcounter{figure}{0}
\renewcommand{\thefigure}{S\arabic{figure}}%
}
\beginsupplement

\bibliographystyle{plain}
\bibliography{paper}

\clearpage
\begin{figure*}
\section*{SUPPLEMENTARY INFORMATION FOR: IGLOO: Integrating global and local biological network alignment}

\textbf{Lei Meng\,$^{1,2,3,4}$, Joseph Crawford\,$^{1,2,3}$, Aaron Striegel\,$^{1,4}$, and Tijana Milenkovi\'{c}\,$^{1,2,3,*}$}\\
$^{1}$Department of Computer Science and Engineering\\
$^{2}$ECK Institute of Global Health \\
$^{3}$Interdisciplinary Center for Network Science and Applications (iCeNSA)\\
$^{4}$Wireless Institute\\
University of Notre Dame, Notre Dame, IN 46556\\
$^*$To whom correspondence should be addressed

\end{figure*}
\begin{figure*}
\centering
\includegraphics[width=0.48\textwidth]{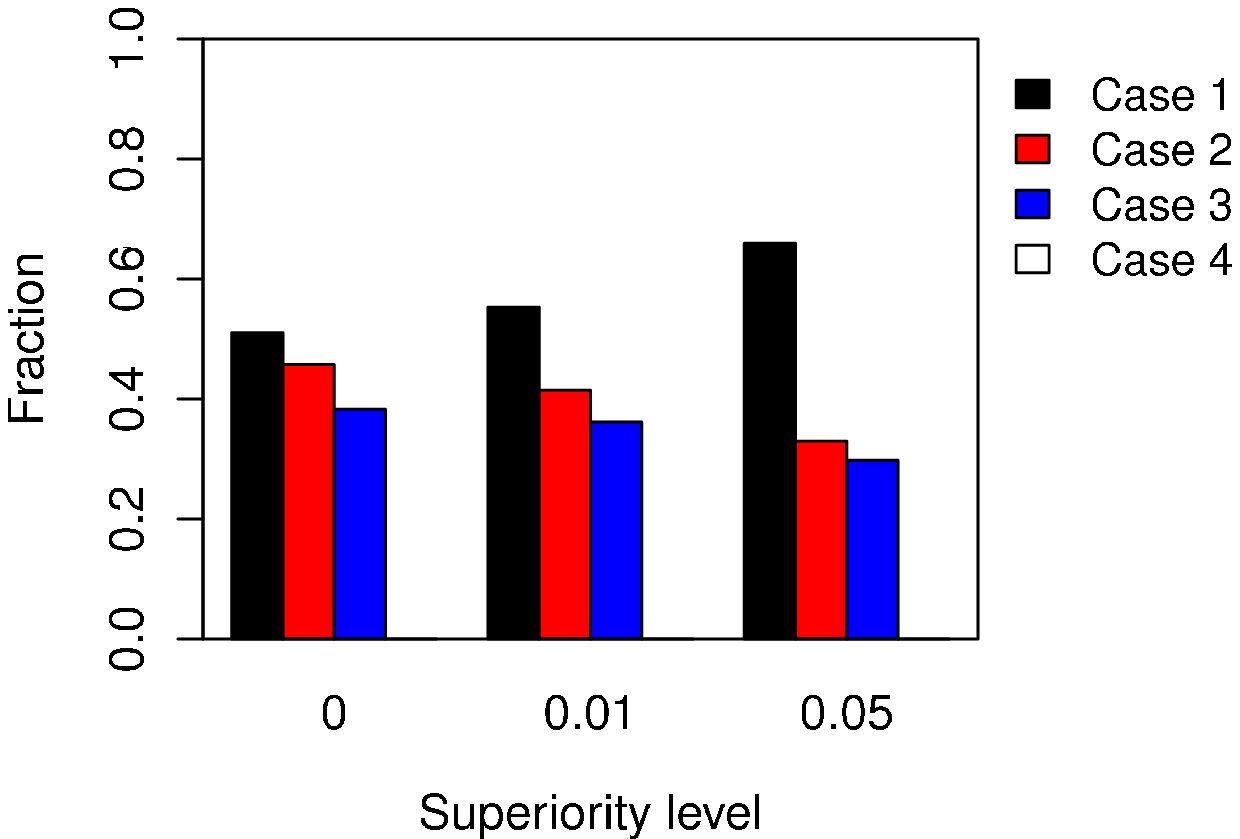}\\
\hspace{-.6cm}\textbf{(a)} LNA and GNA combined

\includegraphics[width=0.48\textwidth]{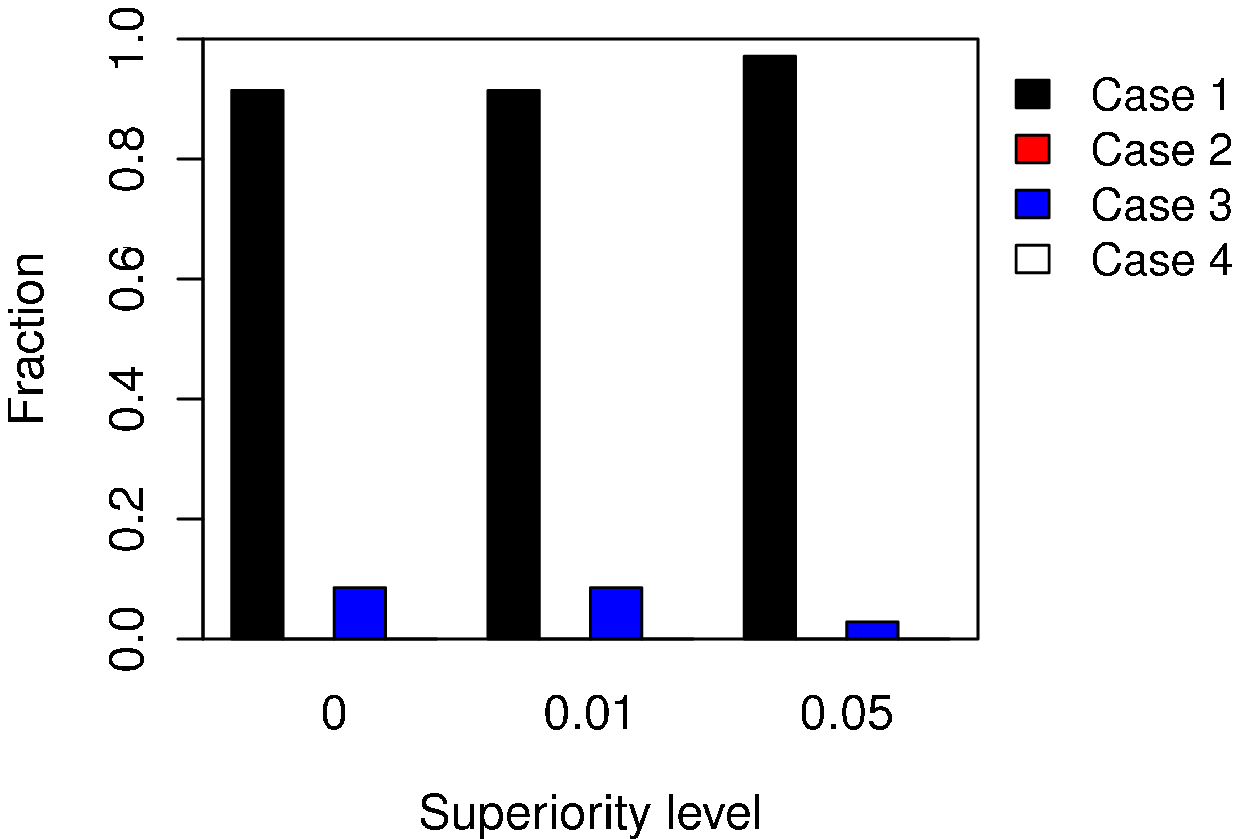}
\includegraphics[width=0.48\textwidth]{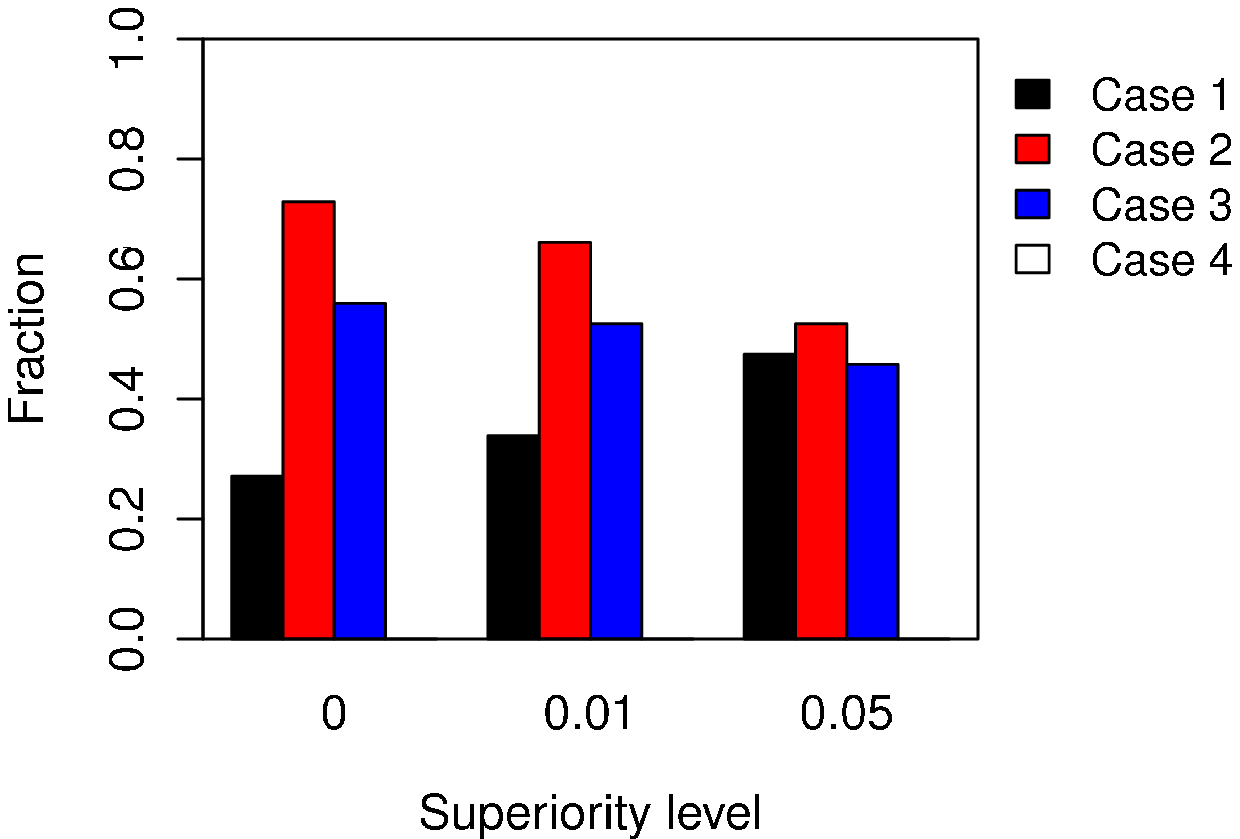}\\
\hspace{-0.5cm}\textbf{(b)} LNA \hspace{7.5cm}\textbf{(c)} GNA

\caption{Overall comparison of IGLOO (the best of its versions) and \textbf{(a)} LNA and GNA combined, \textbf{(b)} LNA, and \textbf{(c)} GNA, when considering five different versions of IGLOO (IGLOO 0-4) under AlignMCL in the first step of the algorithm. The comparison is shown for three different method ``superiority levels'' (denoted as $p$): 0\%, 1\%, and 5\%. By a ``superiority level'', we mean the following. Given two methods $A$ and $B$ with alignment quality scores $x$ and $y$, respectively, if $\frac{|x-y|}{max(x, y)} \leq p$, we say that $A$ and $B$ are comparable; otherwise, if $x$ is greater/less than $y$, we say that $A$ is superior/inferior to $B$. For a given network pair and a given existing method, only the best version of IGLOO is considered. The four cases are as follows: 1) IGLOO is comparable or superior both topologically and functionally; 2) IGLOO is comparable or superior only functionally but not topologically; 3) IGLOO is comparable or superior only topologically but not functionally; and 4) IGLOO is inferior both topologically and functionally. The \emph{y}-axes indicate the percentage of the combinations of the existing NA methods and networks pairs for which the given case occurs.}
\label{fig:comparison_alignmcl}
\end{figure*}

\begin{figure*}
\centering

\includegraphics[width=0.48\textwidth]{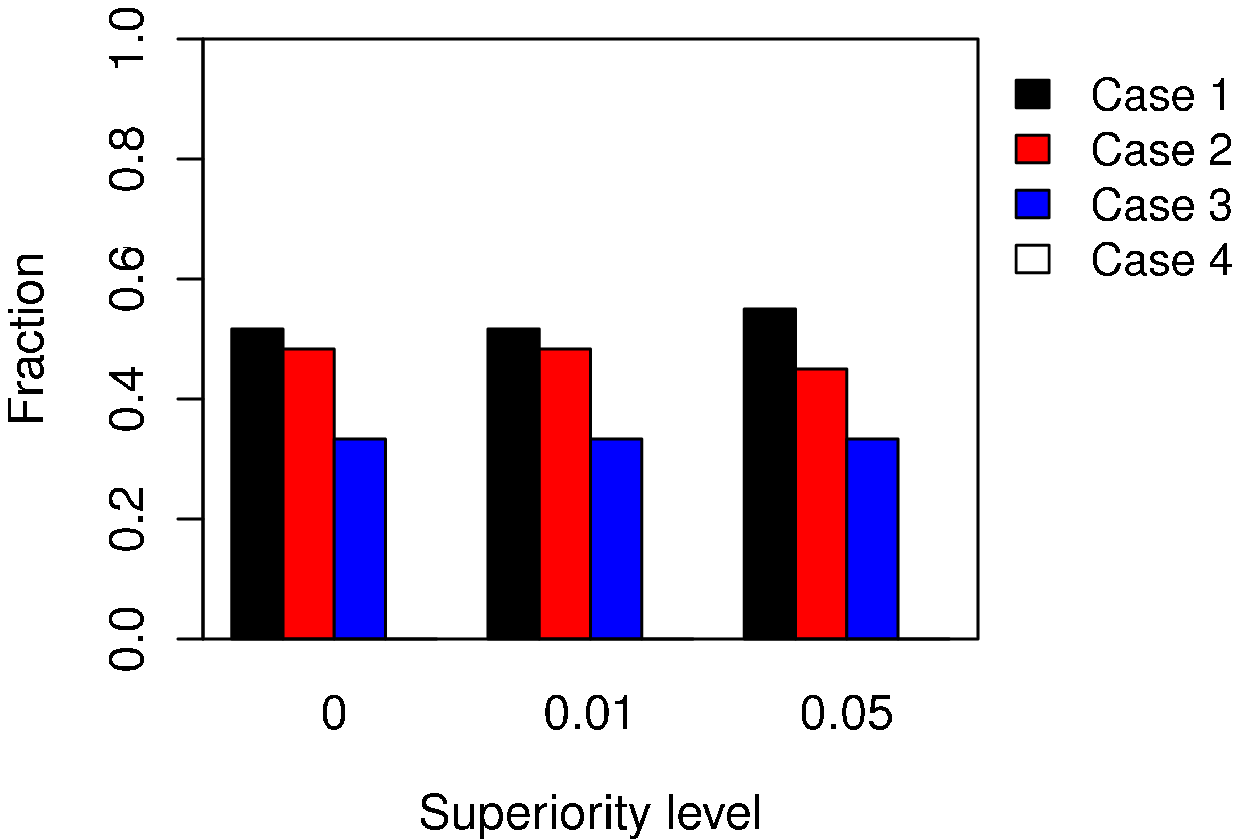}\\
\hspace{-.6cm}\textbf{(a)} LNA and GNA combined

\includegraphics[width=0.48\textwidth]{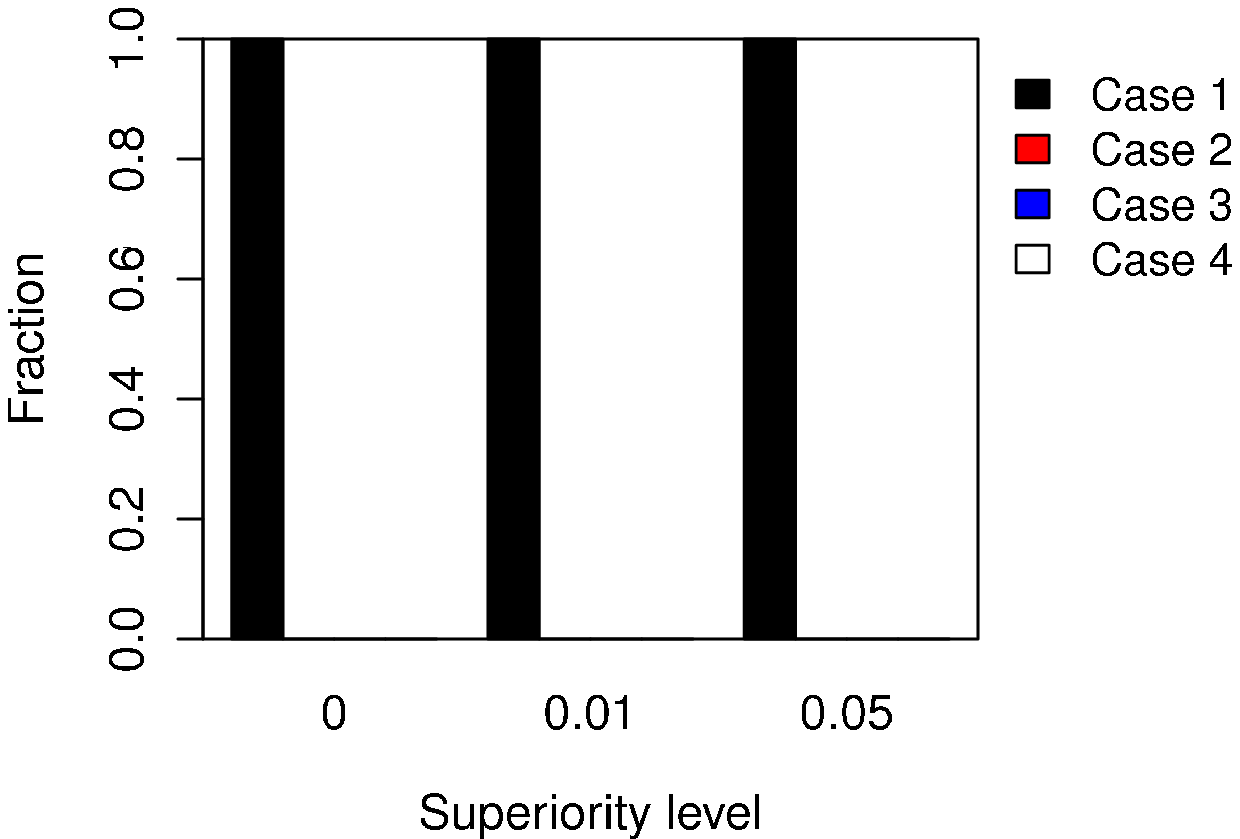}
\includegraphics[width=0.48\textwidth]{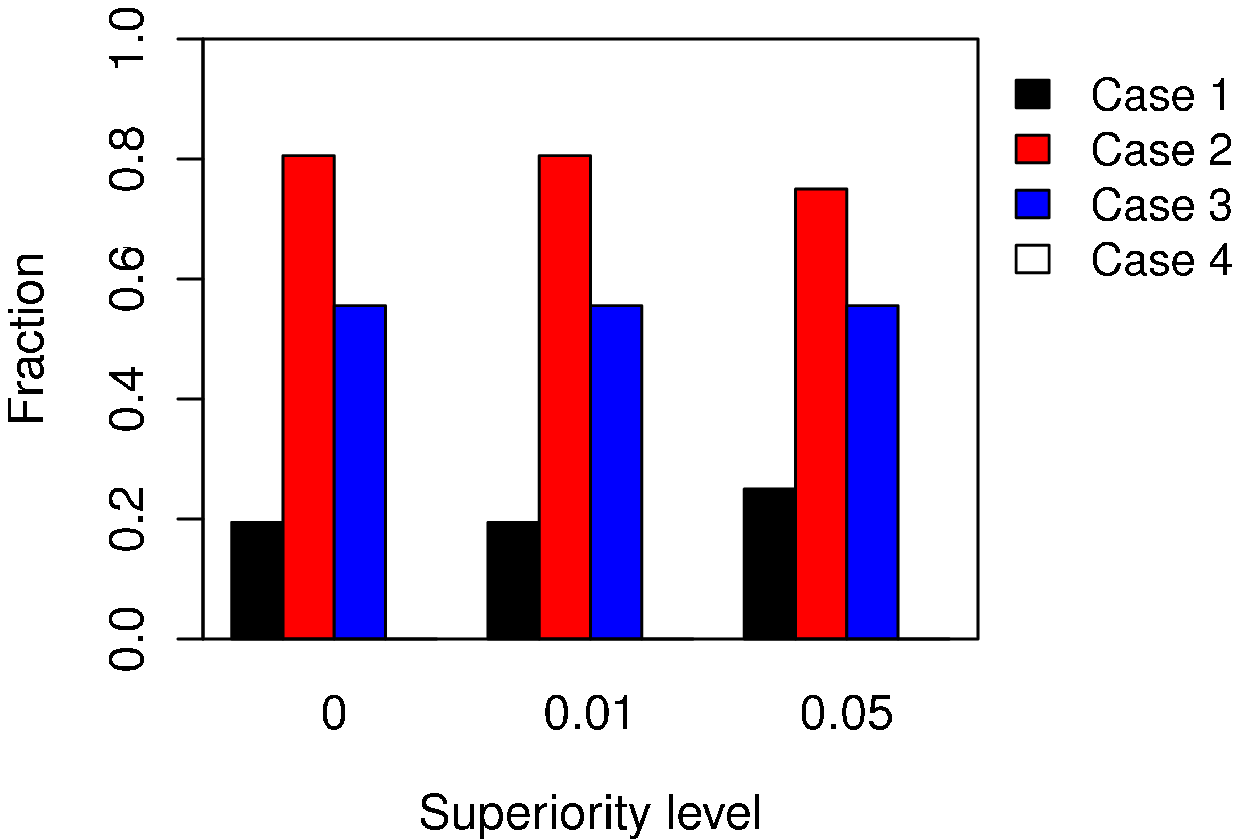}\\
\hspace{-0.5cm}\textbf{(b)} LNA \hspace{7.5cm}\textbf{(c)} GNA

\caption{Overall comparison of IGLOO (the best of its versions) and \textbf{(a)} LNA and GNA combined, \textbf{(b)} LNA, and \textbf{(c)} GNA, when considering five different versions of IGLOO (IGLOO 0-4) under AlignNemo in the first step of the algorithm. The comparison is shown for three different method ``superiority levels'' (denoted as $p$): 0\%, 1\%, and 5\%. By a ``superiority level'', we mean the following. Given two methods $A$ and $B$ with alignment quality scores $x$ and $y$, respectively, if $\frac{|x-y|}{max(x, y)} \leq p$, we say that $A$ and $B$ are comparable; otherwise, if $x$ is greater/less than $y$, we say that $A$ is superior/inferior to $B$. For a given network pair and a given existing method, only the best version of IGLOO is considered. The four cases are as follows: 1) IGLOO is comparable or superior both topologically and functionally; 2) IGLOO is comparable or superior only functionally but not topologically; 3) IGLOO is comparable or superior only topologically but not functionally; and 4) IGLOO is inferior both topologically and functionally. The \emph{y}-axes indicate the percentage of the combinations of the existing NA methods and networks pairs for which the given case occurs.}
\label{fig:comparison_alignnemo}
\end{figure*}

\begin{figure*}
\centering

\begin{minipage}{0.48\textwidth}
\centering
yeast-fly (Y2H$_1$)\hspace{1cm}
\end{minipage}
\begin{minipage}{0.48\textwidth}
\centering
yeast-worm (Y2H$_1$)\hspace{1cm}
\end{minipage}
\vspace{-.2in}

\begin{minipage}{0.48\textwidth}
\centering
\includegraphics[width=\textwidth]{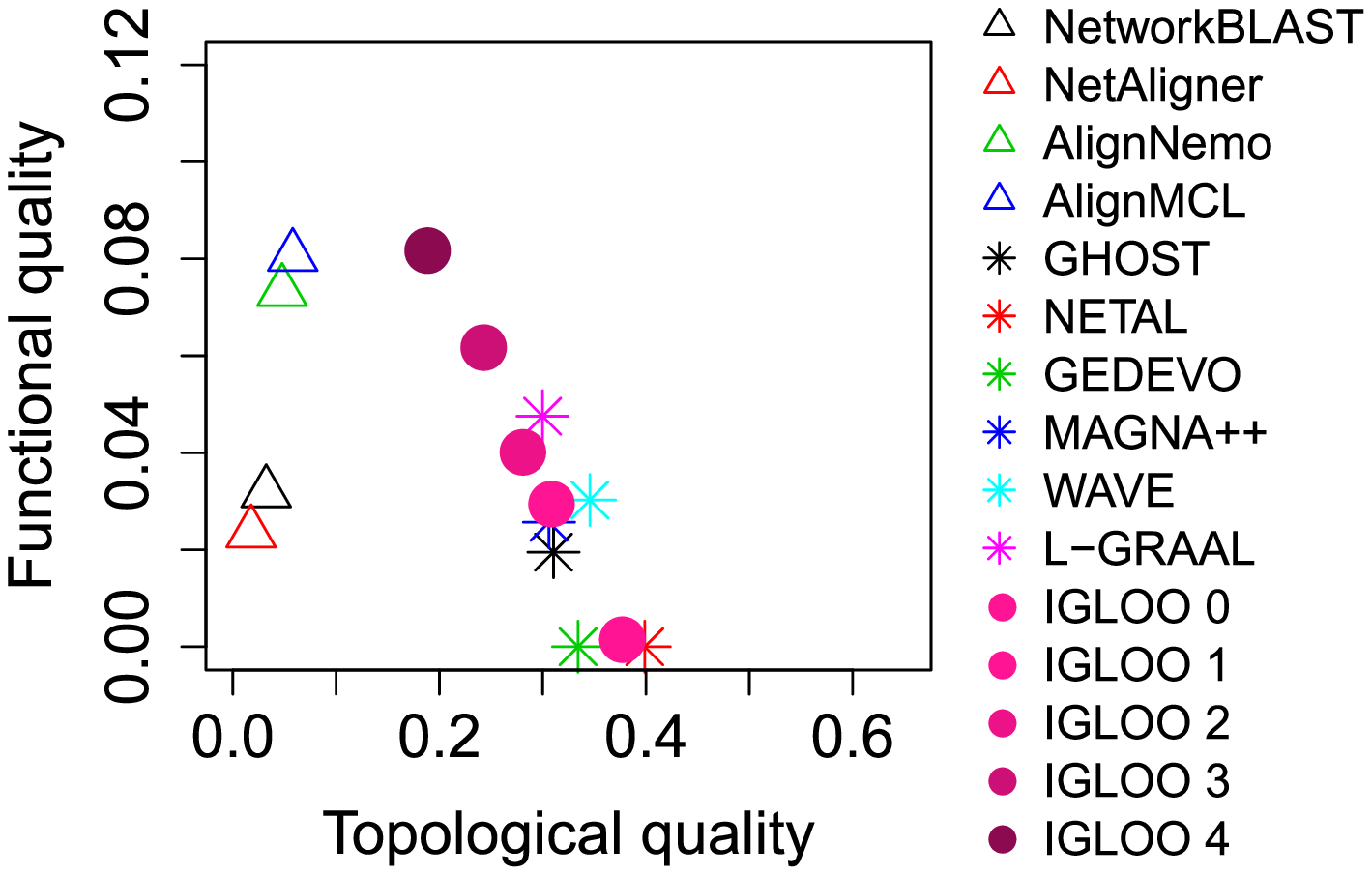}
\end{minipage}
\begin{minipage}{0.48\textwidth}
\centering
\includegraphics[width=\textwidth]{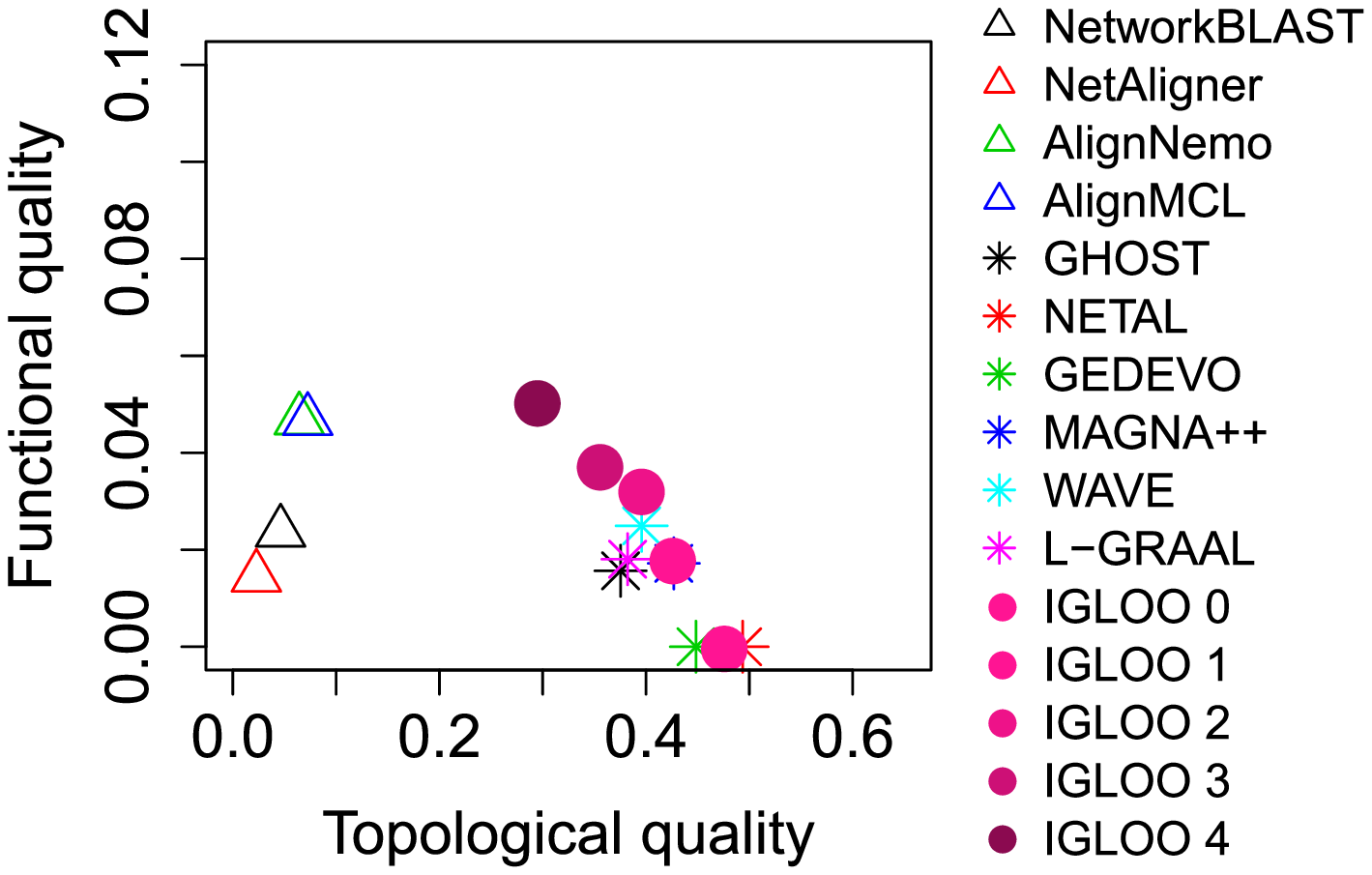}
\end{minipage}

\begin{minipage}{0.48\textwidth}
\centering
worm-fly (Y2H$_1$)\hspace{1cm}
\end{minipage}
\begin{minipage}{0.48\textwidth}
\centering
yeast-human (Y2H$_2$)\hspace{1cm}
\end{minipage}
\vspace{-.2in}

\begin{minipage}{0.48\textwidth}
\centering
\includegraphics[width=\textwidth]{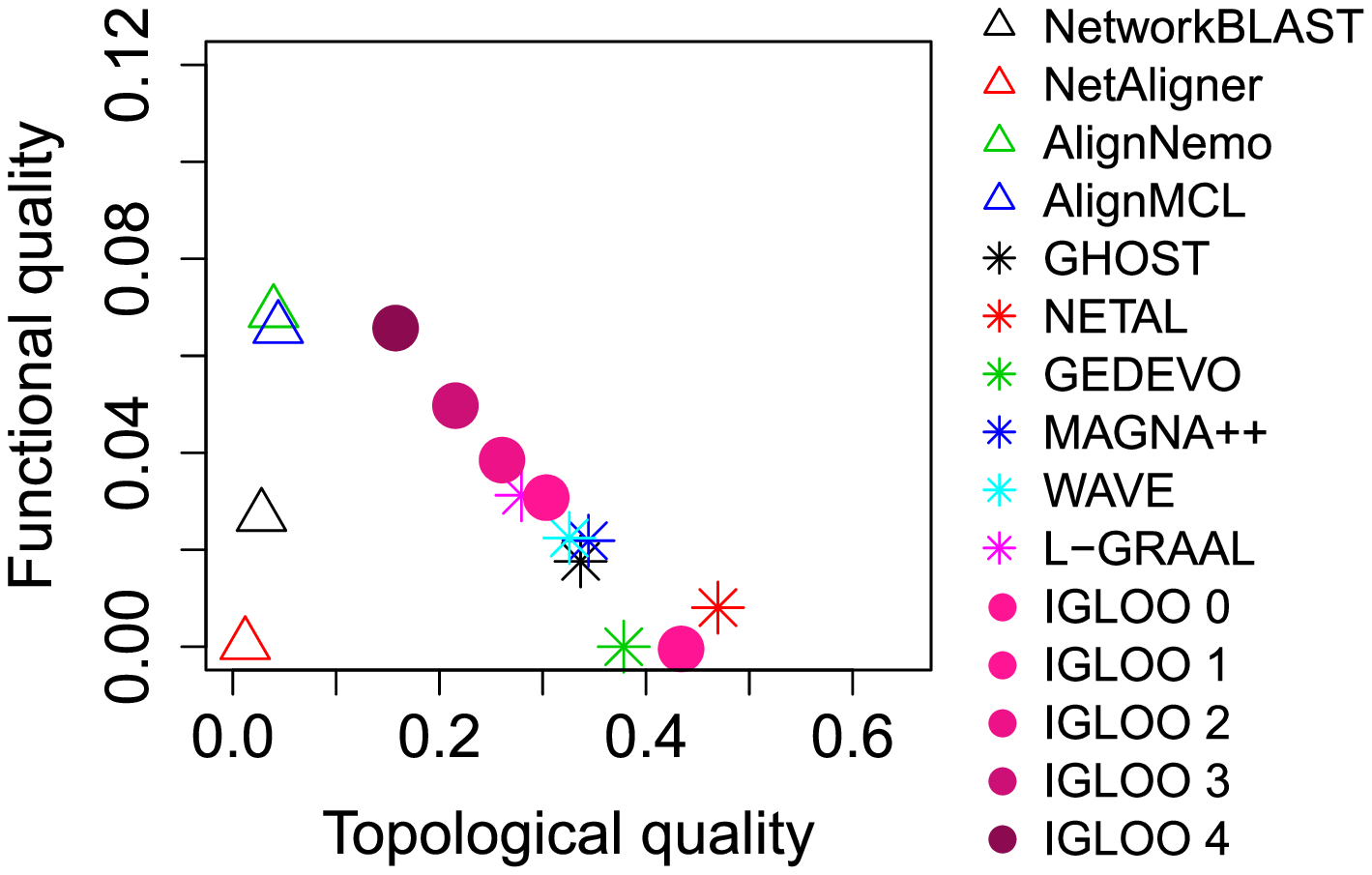}
\end{minipage}
\begin{minipage}{0.48\textwidth}
\centering
\includegraphics[width=\textwidth]{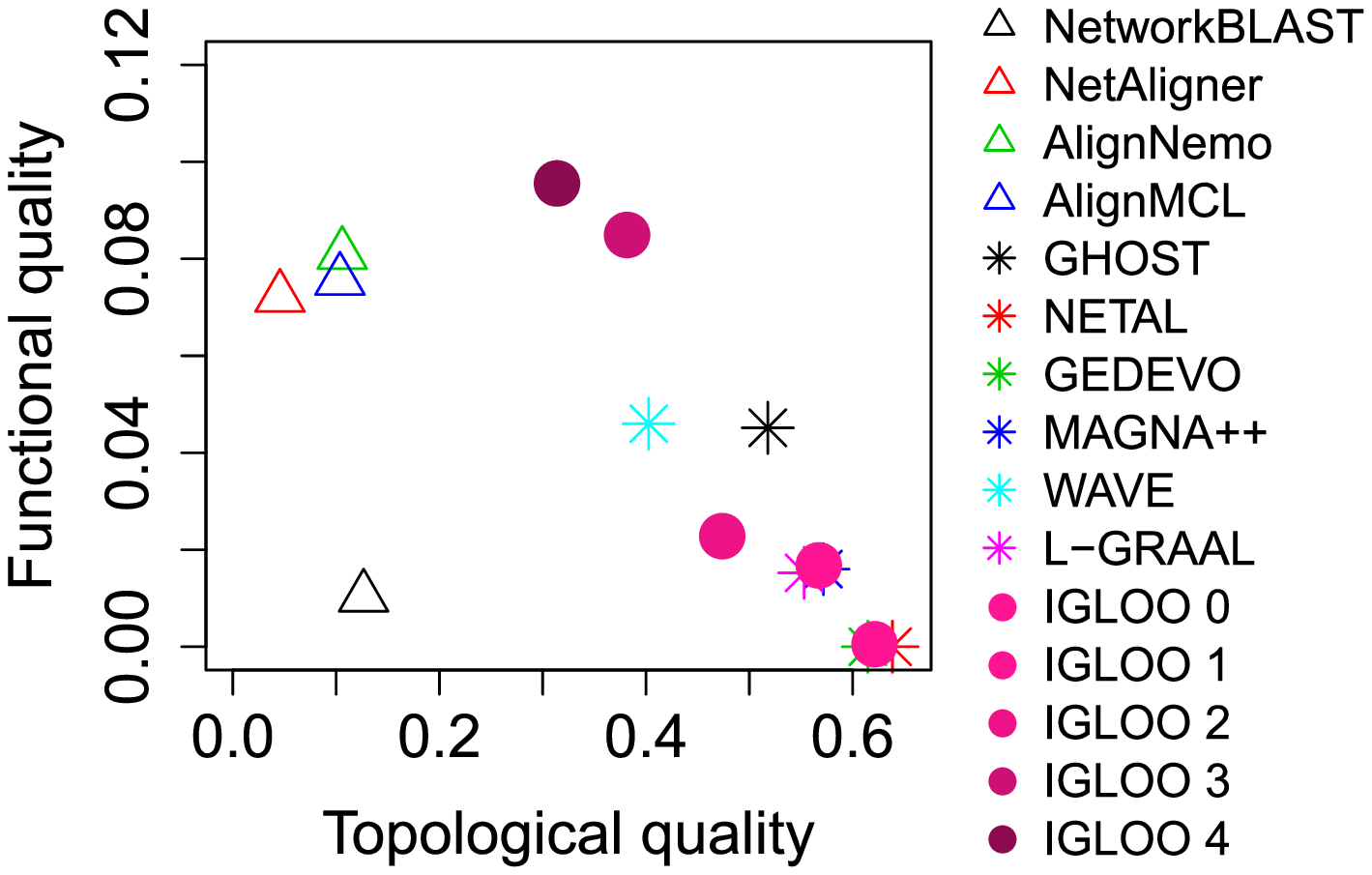}
\end{minipage}

\begin{minipage}{0.48\textwidth}
\centering
yeast-worm (PHY$_1$)\hspace{1cm}
\end{minipage}
\begin{minipage}{0.48\textwidth}
\centering
fly-worm (PHY$_1$)\hspace{1cm}
\end{minipage}
\vspace{-.2in}

\begin{minipage}{0.48\textwidth}
\centering
\includegraphics[width=\textwidth]{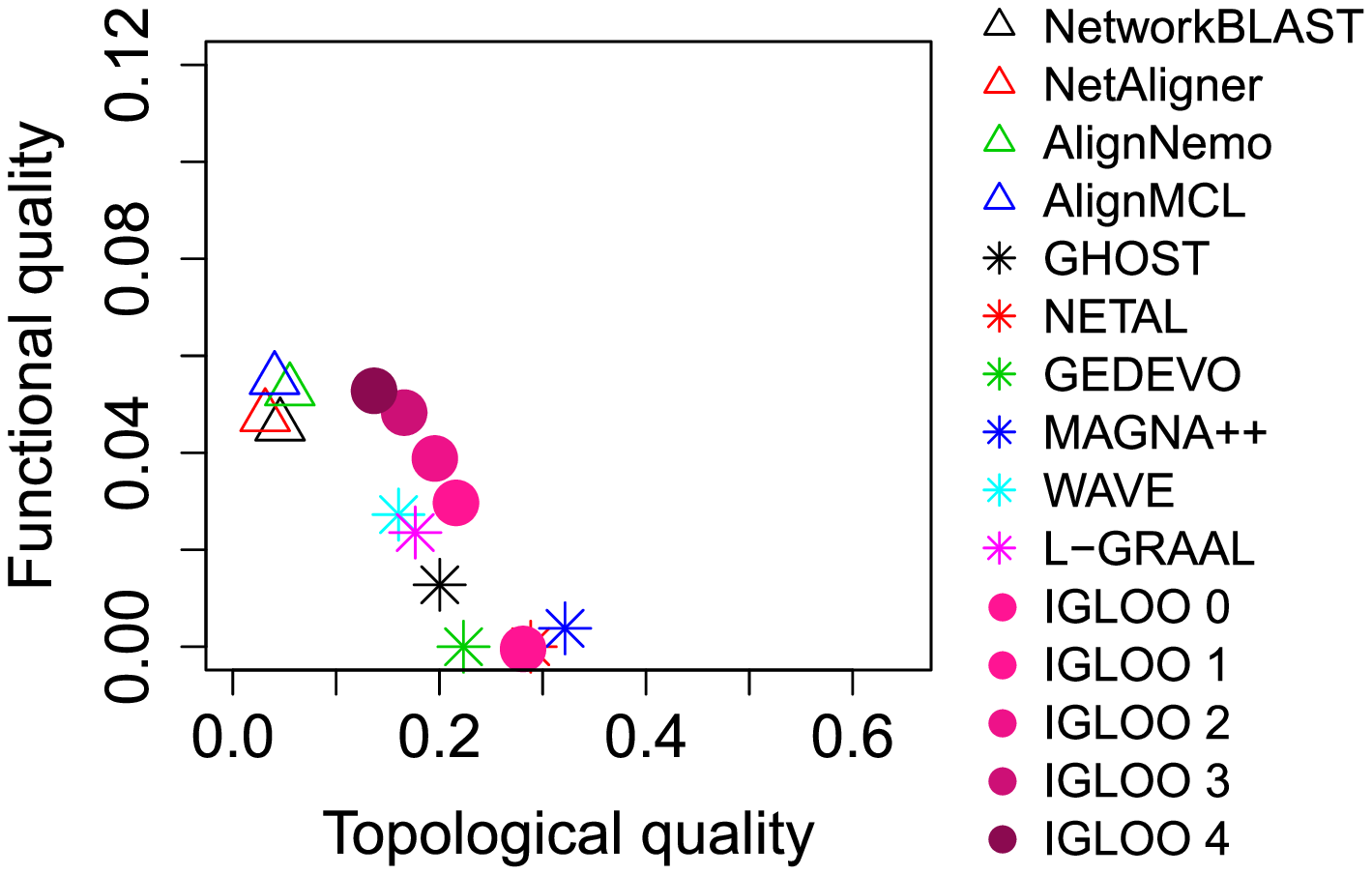}
\end{minipage}
\begin{minipage}{0.48\textwidth}
\centering
\includegraphics[width=\textwidth]{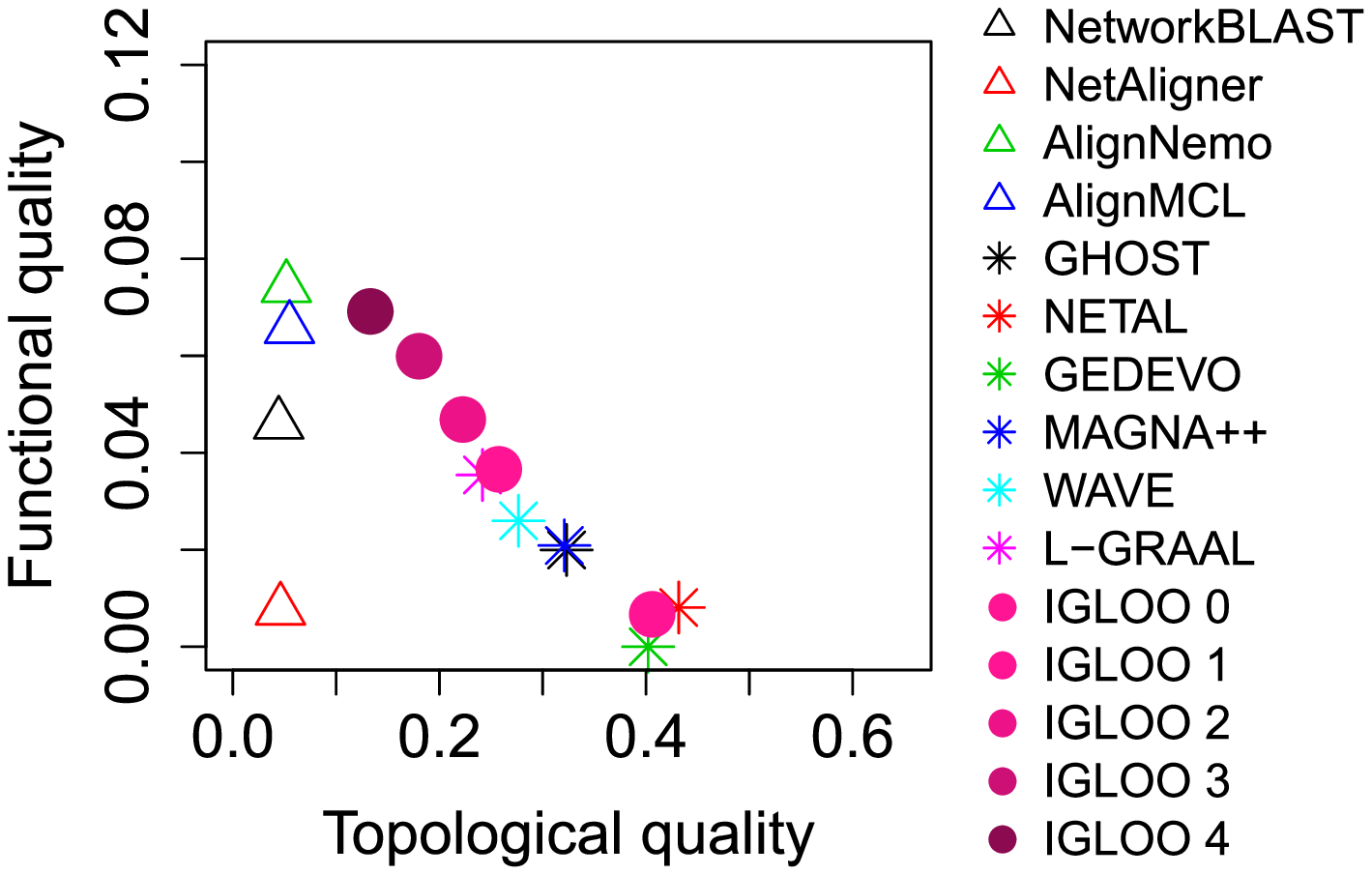}
\end{minipage}

\caption{Topological (NCV-GS$^3$; $x$-axis) and functional (F-PF; $y$-axis) alignment quality for the existing LNA methods (triangles), existing GNA methods (stars) and IGLOO versions (circles), for each aligned network pair, when considering AlignMCL in the first step of IGLOO algorithm.}
\label{fig:top_bio_alignmcl1}
\end{figure*}

\begin{figure*}
\centering

\begin{minipage}{0.48\textwidth}
\centering
yeast-fly (Y2H$_1$)\hspace{1cm}
\end{minipage}
\begin{minipage}{0.48\textwidth}
\centering
yeast-worm (Y2H$_1$)\hspace{1cm}
\end{minipage}
\vspace{-.2in}

\begin{minipage}{0.48\textwidth}
\centering
\includegraphics[width=\textwidth]{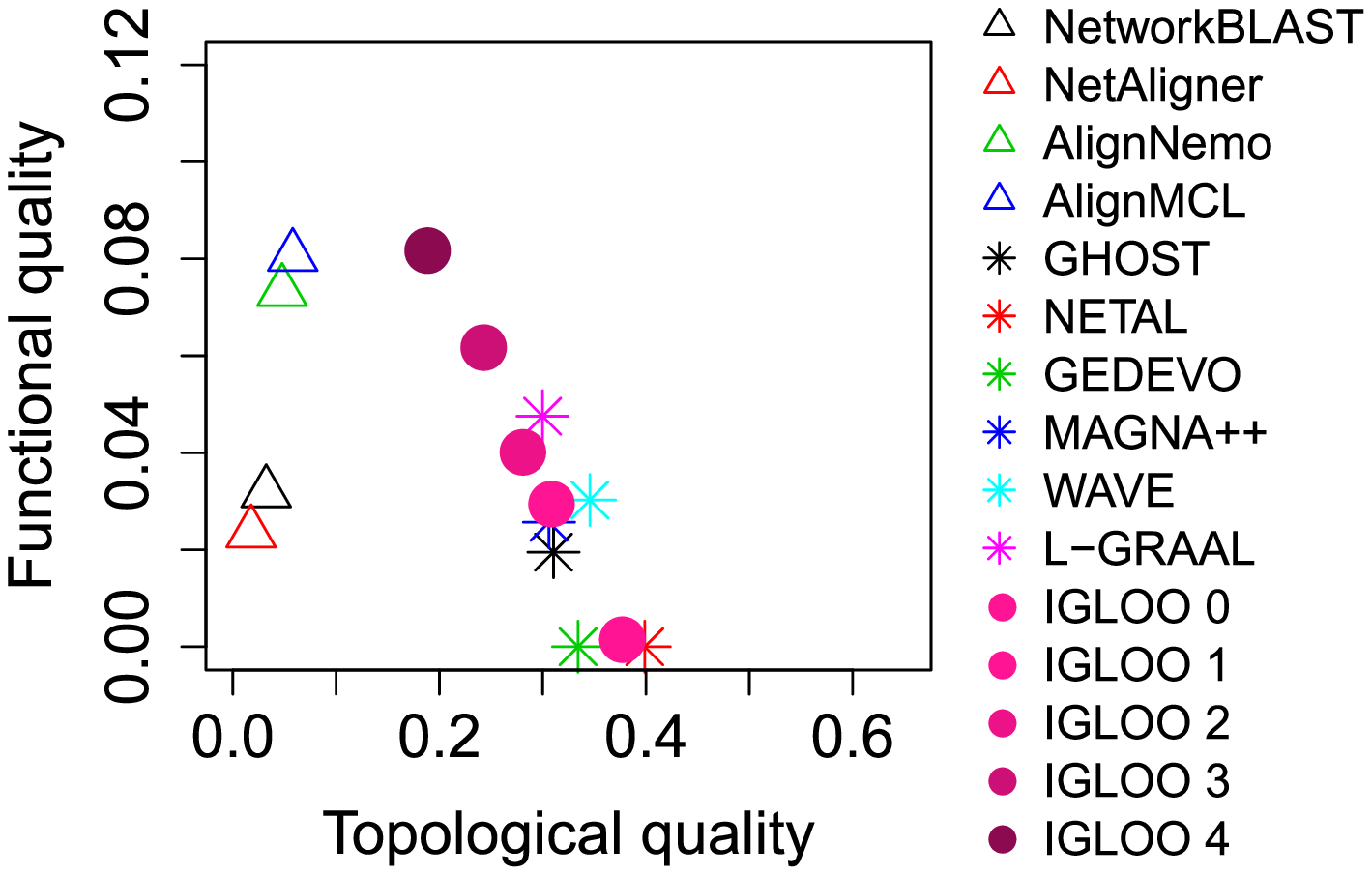}
\end{minipage}
\begin{minipage}{0.48\textwidth}
\centering
\includegraphics[width=\textwidth]{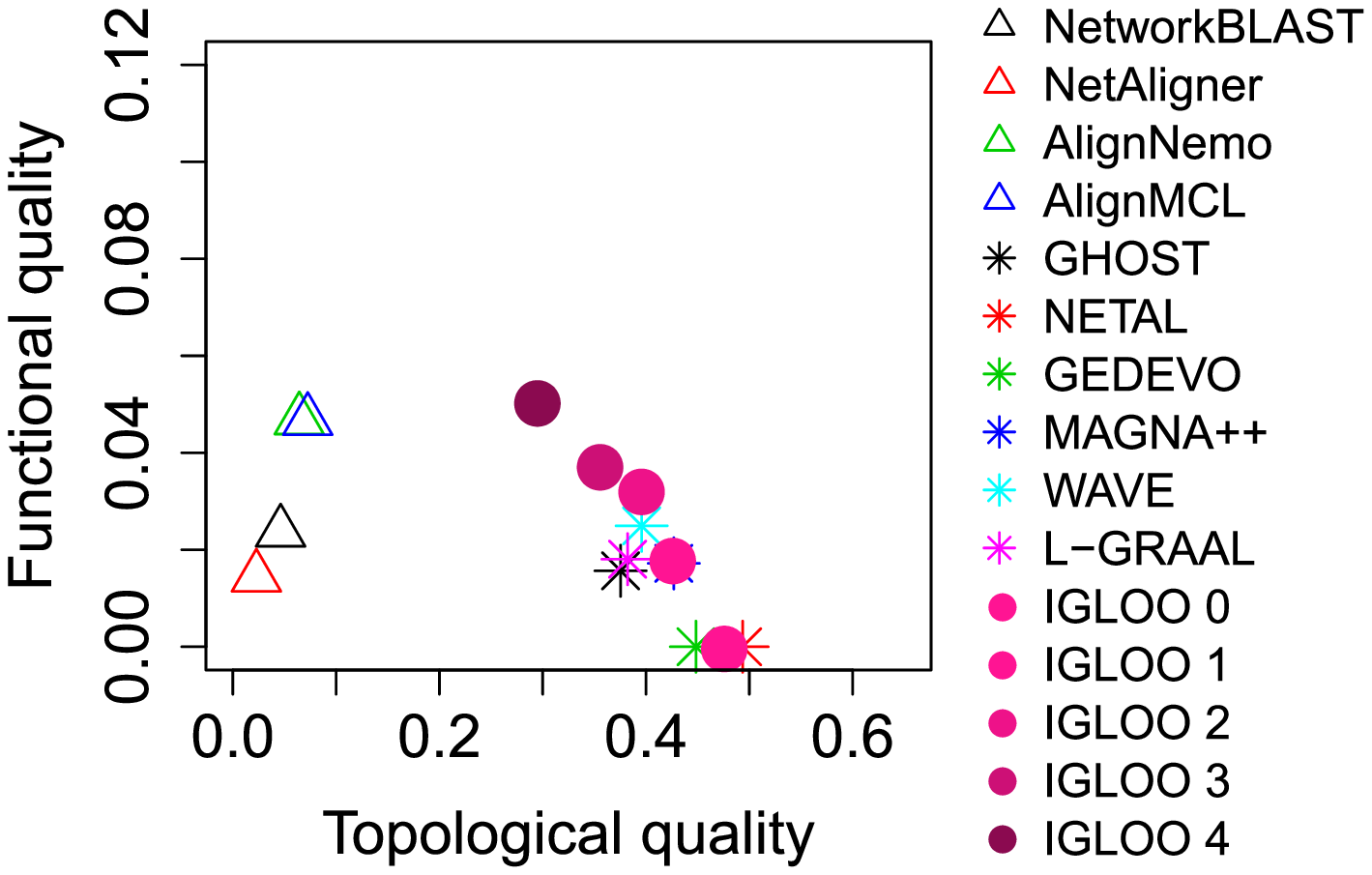}
\end{minipage}

\begin{minipage}{0.48\textwidth}
\centering
worm-fly (Y2H$_1$)\hspace{1cm}
\end{minipage}
\begin{minipage}{0.48\textwidth}
\centering
yeast-human (Y2H$_2$)\hspace{1cm}
\end{minipage}
\vspace{-.2in}

\begin{minipage}{0.48\textwidth}
\centering
\includegraphics[width=\textwidth]{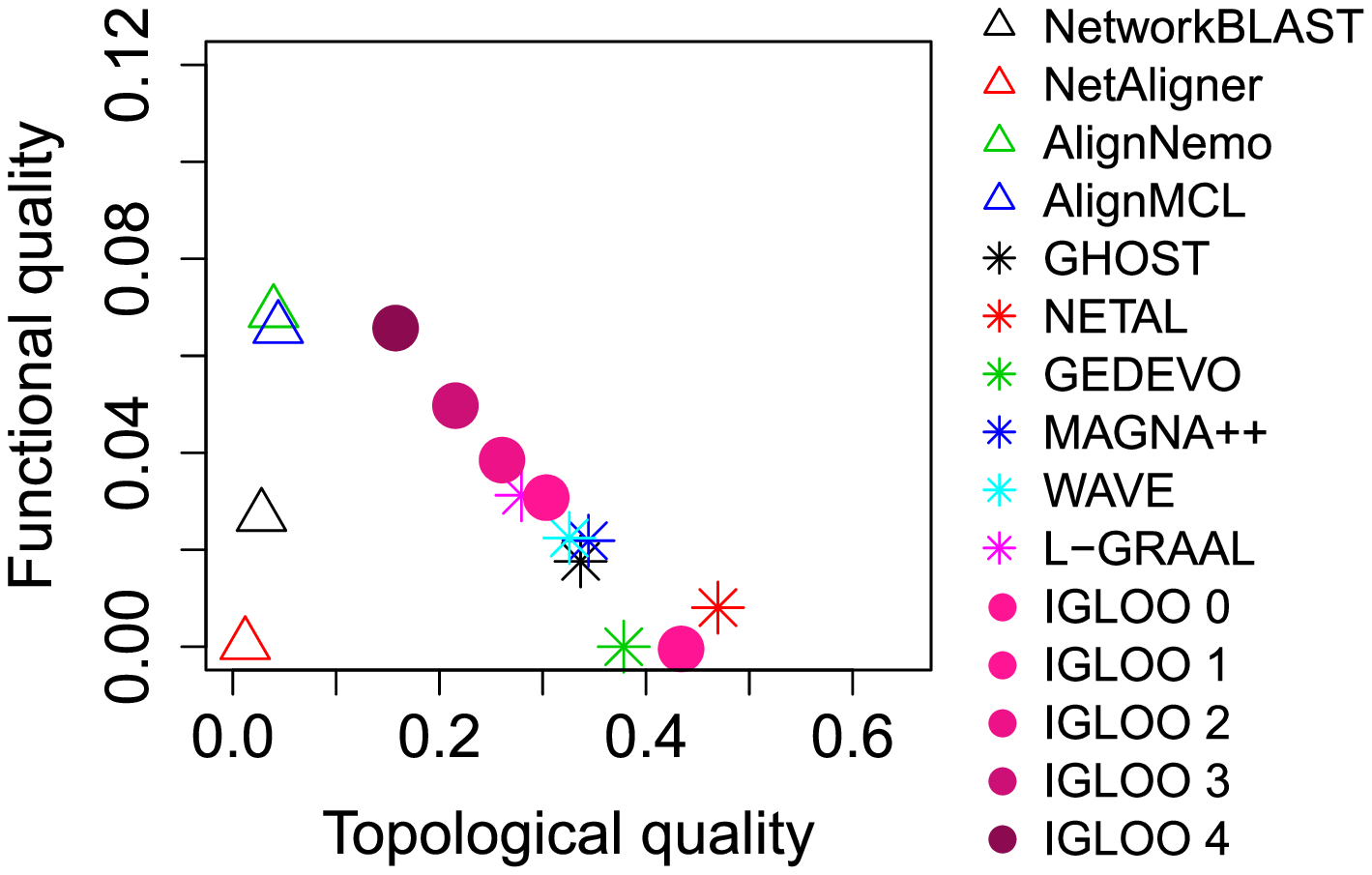}
\end{minipage}
\begin{minipage}{0.48\textwidth}
\centering
\includegraphics[width=\textwidth]{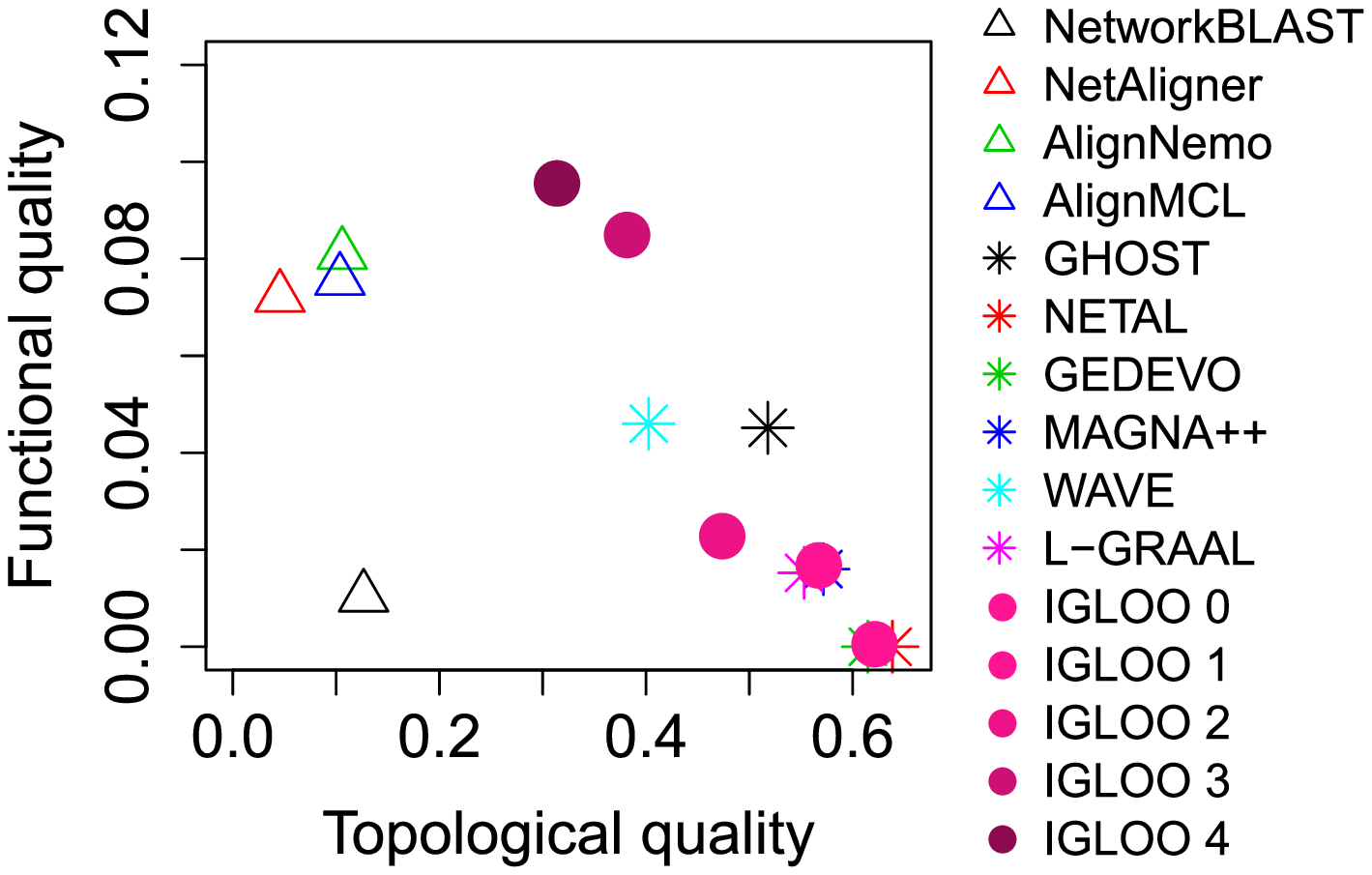}
\end{minipage}

\begin{minipage}{0.48\textwidth}
\centering
yeast-worm (PHY$_1$)\hspace{1cm}
\end{minipage}
\begin{minipage}{0.48\textwidth}
\centering
fly-worm (PHY$_1$)\hspace{1cm}
\end{minipage}
\vspace{-.2in}

\begin{minipage}{0.48\textwidth}
\centering
\includegraphics[width=\textwidth]{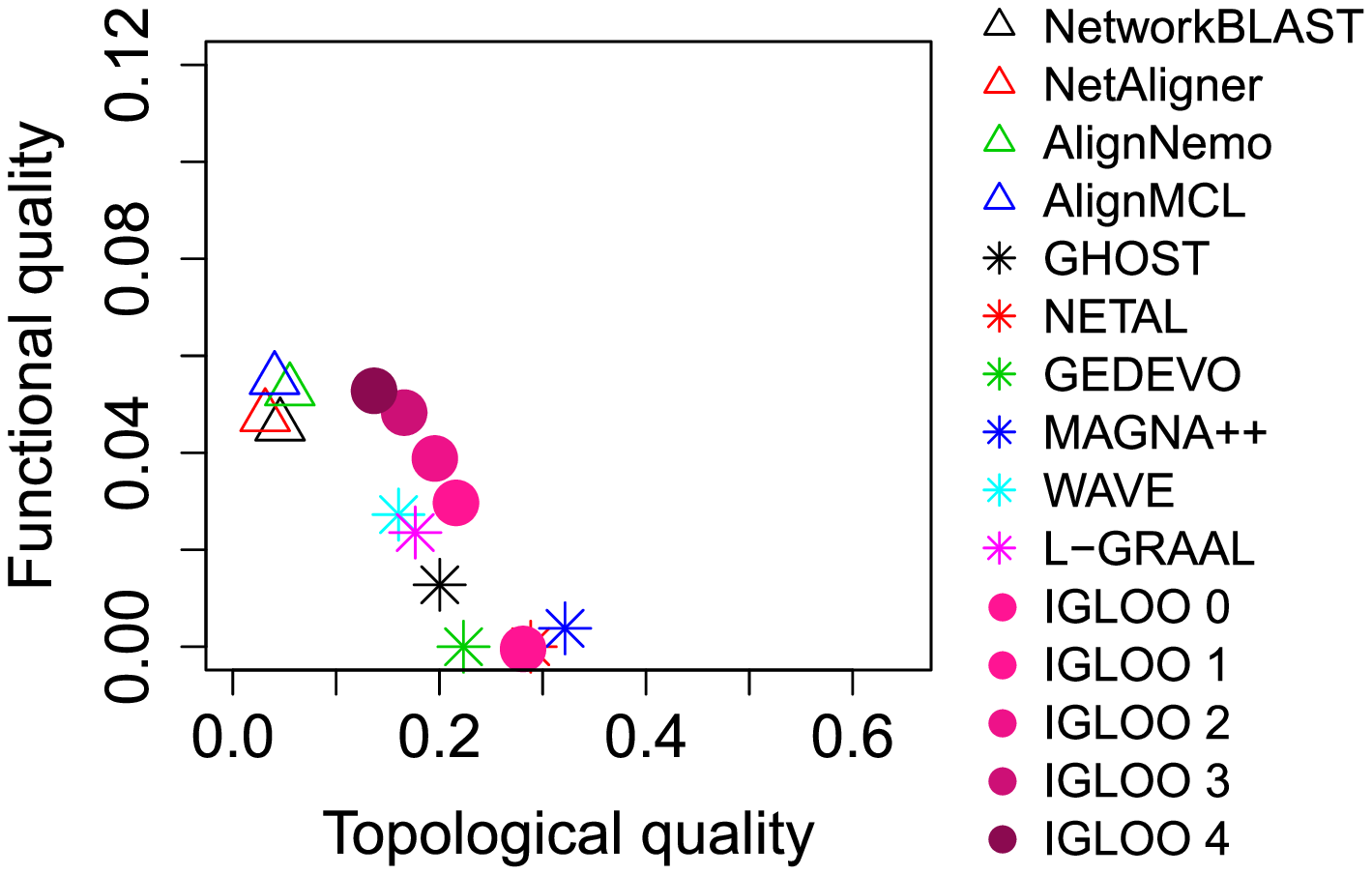}
\end{minipage}
\begin{minipage}{0.48\textwidth}
\centering
\includegraphics[width=\textwidth]{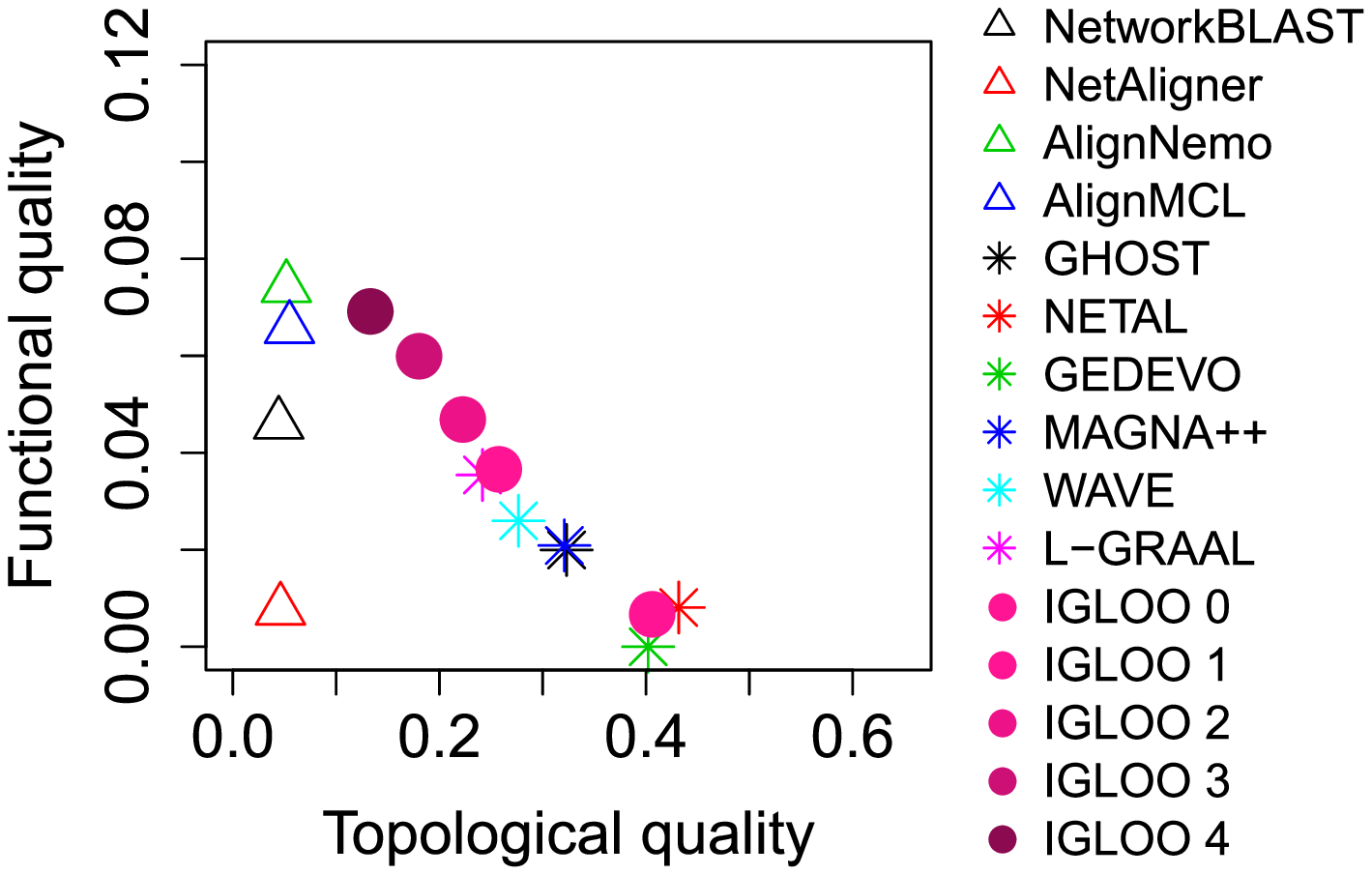}
\end{minipage}
\caption{Topological (NCV-GS$^3$; $x$-axis) and functional (F-PF; $y$-axis) alignment quality for the existing LNA methods (triangles), existing GNA methods (stars) and IGLOO versions (circles), for each aligned network pair, when considering AlignNemo in the first step of IGLOO algorithm.}
\label{fig:top_bio_alignnemo}
\end{figure*}

\end{document}